\documentclass[10pt,twocolumn,twoside]{IEEEtran}
\usepackage{bm}
\usepackage{amsmath}
\usepackage{amsthm}
\usepackage{graphicx}
\usepackage{amssymb}
\usepackage{epsfig}
\usepackage{epstopdf}
\usepackage{float}
\usepackage{mathrsfs}
\usepackage{caption}
\usepackage{subcaption}
\usepackage{longtable}

\usepackage{xtab}
\usepackage{color}
\usepackage{ifthen}
\usepackage{cite}

\usepackage{setspace}

\newtheorem{rem}{Remark}

\newtheorem{theo}{Theorem}

\newtheorem{ass}{Assumption}
\newtheorem{prop}{Proposition}

\begin{document}

\allowdisplaybreaks

\title{Decentralized and Distributed Temperature Control via HVAC Systems in Energy Efficient Buildings}
\author{Xuan Zhang, Wenbo Shi, Bin Yan, Ali Malkawi and Na Li
\thanks{Some of the material in this paper appeared in the Proceedings of the 2016 IEEE Global Conference on Signal and Information Processing, December 7-9, 2016, Arlington, VA, USA.}
\thanks{This work was supported by NSF ECCS 1548204, 1608509, NSF CAREER 1553407, and Harvard Center for Green Buildings and Cities. X. Zhang, W. Shi, B. Yan and A. Malkawi are with Harvard Center for Green Buildings and Cities, 20 Sumner Road, Cambridge, MA 02138, USA (email: xuan\_zhang@g.harvard.edu, \{wshi, byan, amalkawi\}@gsd.harvard.edu). X. Zhang and N. Li are with the School of Engineering and Applied Sciences, Harvard University, 29 Oxford Street, Cambridge, MA 02138, USA (email: nali@seas.harvard.edu).} %\textit{Corresponding Author: N. Li.}
}

\maketitle

\begin{abstract}
In this paper, we design real-time decentralized and distributed control schemes for Heating Ventilation and Air Conditioning (HVAC) systems in energy efficient buildings. The control schemes balance user comfort and energy saving, and are implemented without measuring or predicting exogenous disturbances. Firstly, we introduce a thermal dynamic model of building systems and formulate a steady-state resource allocation problem, which aims to minimize the aggregate deviation between zone temperatures and their set points, as well as the building energy consumption, subject to practical operating constraints, by adjusting zone flow rates. Because this problem is nonconvex, we propose two methods to (approximately) solve it and to design the real-time control. In the first method, we present a convex relaxation approach to solve an approximate version of the steady-state optimization problem, where the heat transfer between neighboring zones is ignored. We prove the tightness of the relaxation and develop a real-time decentralized algorithm to regulate the zone flow rate. In the second method, we introduce a mild assumption under which the original optimization problem becomes convex, and then a real-time distributed algorithm is developed to regulate the zone flow rate. In both cases, the thermal dynamics can be driven to equilibria which are optimal solutions to those associated steady-state optimization problems. Finally, numerical examples are provided to illustrate the effectiveness of the designed control schemes.
\end{abstract}


\begin{spacing}{0.96}

\section{Introduction}
It is reported that buildings are responsible for $40\%$ of energy consumption, $70\%$ of electricity consumption, and result in $30\%$ of greenhouse gas emission~\cite{UNEP09}. Roughly speaking, Heating Ventilation and Air Conditioning (HVAC) systems in buildings account for $40\%$ of the energy use~\cite{USE12}. Therefore, making HVAC systems more energy efficient is urgent for improving environmental sustainability.

To date, various control techniques have been developed for HVAC systems, including gain scheduling, optimal control, robust control, nonlinear adaptive control, Model Predictive Control (MPC), intelligent control based on artificial neural network, fuzzy logic, and genetic algorithm, and so forth~\cite{NaiR11i,NaiR11ii,OldPJ12,AswMT12,MaKDB12}. Compared with the conventional on/off plus Proportional-Integral-Derivative (PID) control, these techniques are more robust and energy efficient. However, most of them requires centralized operation with heavy burdens of sensing, communication and computation, leading to much higher implementation cost than the conventional on/off plus PID control.

On the other hand, smart sensing, communication, computing, and actuation technologies have been stimulating the emergence of distributed/decentralized control in network systems, including the smart grid~\cite{Far10,ZhaP15,ZhaKMP15}, smart buildings~\cite{ShiLXCG14,ShiLC17}, mobile robots~\cite{JadLM03}, and intelligent transportation systems~\cite{Wan10}. The advantages of distributed/decentralized control include: good scalability as the network grows; reduction in measurement, communication and computation compared with centralized control; privacy preserving. Thus, applying distributed/decentralized control to HVAC system control and optimization becomes an area of active research. Representative work includes, for example, distributed MPC~\cite{MorBDB10,MaAB11,HaoLKS15}, as well as mean-field game based distributed control~\cite{DenBM12}. Among these popular control mechanisms, distributed MPC is the most promising one, especially for large buildings. However, it still requires a large amount of sensing, communication and computation. And in most cases, it needs good prediction of future disturbances, i.e., outdoor temperature, sunlight, indoor occupancy, etc., which may be hard to obtain.

\textbf{Contribution of this paper.} This paper aims to develop real-time control schemes for HVAC systems in commercial buildings. Specifically, we aim to design decentralized and distributed algorithms to guide each thermal zone to properly adapt their supply air flow rates such that system-wide objectives are achieved under given operating conditions. The proposed control schemes (i) are scalable with respect to building structures, (ii) satisfy the system operating constraints, and (iii) ensure system efficiency, reliability and user comfort. Different from literature~\cite{MorBDB10,MaAB11,HaoLKS15,DenBM12}, the control schemes designed in this paper are based on solving steady-state resource allocation problems via gradient algorithms -- they are dynamic feedback controllers that can be implemented without measuring or predicting disturbances, which are different from controllers based on MPC~\cite{OldPJ12,AswMT12,MaKDB12,MorBDB10,MaAB11,HaoLKS15}.

To begin with, we provide the detailed problem setup, including an introduction of the HVAC system configuration, a commonly used thermal dynamic model of the building network, and an optimization problem that aims to minimize a weighted sum of the aggregate deviation between zone temperatures and their set points and the building energy consumption subject to practical operating constraints (Section~\ref{se:setup}). Since the original optimization problem is nonconvex, two different methods are proposed to (approximately) solve it. Firstly, we present a convex relaxation approach in which the heat transfer between neighboring zones is ignored (Section~\ref{se:approximation}). We show that this relaxation is always tight under a mild condition, and develop a decentralized control scheme for real-time zone flow rate regulation. Also, we extend the proposed method to HVAC system management in communities/neighborhoods. Secondly, we introduce a mild assumption under which the original optimization problem is naturally reformulated as a convex one (Section~\ref{se:general}). Then a distributed algorithm is developed for real-time zone flow rate regulation. In both scenarios, the thermal dynamics can be driven to equilibria which are the optimal solutions of those associated steady-state optimization problems. Lastly, two numerical examples are provided to illustrate the effectiveness of the designed control schemes (Section~\ref{se:simulation}), using a building with four adjacent zones.

\noindent\emph{Notation:} $\dot x(t)$ denotes the derivative of a state variable $x(t)$ with respect to time $t$, i.e., $\dot x(t)=\frac{d}{dt}x(t)$. $x\gg(\ll)y$ denotes that $x$ is much greater (less) than $y$. The positive projection of a function $h(y)$ on a variable $x\in[0,+\infty)$, $(h(y))_{x}^{+}$ is:
\begin{gather}
(h(y))_{x}^{+}=\left\{ \begin{array}{ll}
h(y) & \textrm{if $x>0$} \\
\max(0,h(y)) & \textrm{if $x=0$}
\end{array} \right.. \nonumber
\end{gather}

\section{Problem Setup}\label{se:setup}

\subsection{HVAC system in buildings}
\begin{figure}[!t]
\centering
\includegraphics[width=0.45\textwidth]{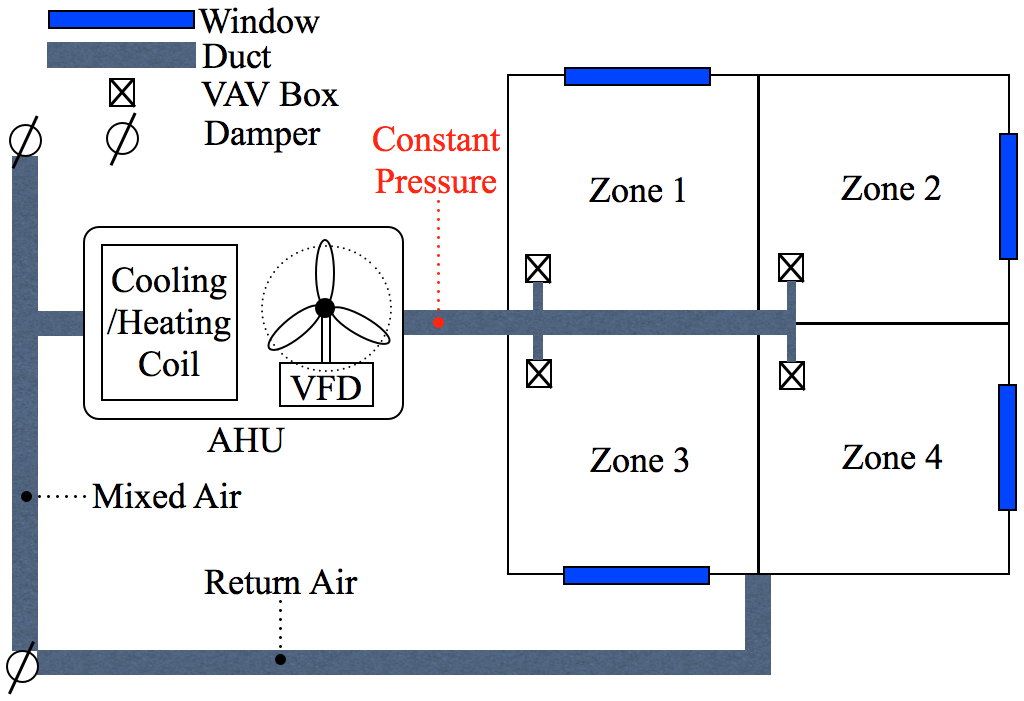}
\caption{Schematic of a typical AHU$\&$VAV system.}
\label{fig:1}
\end{figure}

The schematic of a typical HVAC system is illustrated in Figure~\ref{fig:1}, which consists of an Air Handling Unit (AHU) for the whole building and a set of pressure independent Variable Air Volume (VAV) boxes for each zone~\cite{Sug05}. The AHU is equipped with dampers, a cooling/heating coil, and a Variable Frequency Drive (VFD) fan: the dampers mix the return air from the building with the outside air; the cooling/heating coil cools down/heats up the mixed air; the VFD fan adjusts its own speed based on the total air flow rate/opening controlled by VAV boxes to keep the duct pressure at a certain level, and drives the cooled/heated air to each zone. Each VAV box has a damper to control the air flow rate supplied to the zone, which is considered as the only controllable input to the system in this paper. Also, we will focus on optimizing HVAC operation when the system is in either cooling or heating mode. In the cooling mode, the AHU supply air temperature is often set to $12.8^{\circ}$C which generally provides the required humidity ratio to maintain space thermal comfort~\cite{WanWX07}. In the heating mode, usually dehumidification is not necessary. Therefore, humidity control is not considered in this paper.

%Also, the variables to be controlled here are the air flow rates of VAV boxes under a given supply air temperature. Due to the space limit, the optimization and control of AHU supply air temperature and VAV reheat will be discussed in a future paper. 

\subsection{Thermal dynamic model}
According to the above configuration, we model a building as an undirected connected graph $(\mathcal{N},\mathcal{E})$. Here $\mathcal{N}$ is the set of nodes representing zones/rooms, and $\mathcal{E}\subseteq\mathcal{N}\times\mathcal{N}$ is the set of edges. An edge $(i,j)\in\mathcal{E}$ means that zones $i$ and $j$ are neighbors. Let $\mathcal{N}(i)$ denote the set of neighboring zones of zone $i$. The thermal dynamics for the building is described by a reduced Resistance-Capacitance (RC) model~\cite{LinMB12} (more discussion on the model is available in Remark 1):
\begin{align}\label{equ:thermalmodel}
C_i\dot T_i=\frac{T^o-T_i}{R_i}+\sum_{j\in\mathcal{N}(i)}\frac{T_j-T_i}{R_{ij}}+c_am_i(T^s&-T_i)+Q_i \nonumber\\ &i\in\mathcal{N}
\end{align}
where $C_i$ is the thermal capacitance, $T_i$ is the indoor temperature, $T^o$ is the outdoor temperature, $R_i$ is the thermal resistance of the wall and window separating zone $i$ and outside, $R_{ij}$ is the thermal resistance of the wall separating zones $i$ and $j$, $c_a$ is the specific heat of the air, $m_i$ is the flow rate of the supply air, $T^s$ is the supply air temperature which is usually a constant~\cite{HaoLKS15}, and $Q_i\ge0$ is the heat gain from exogenous sources (e.g., user activity, solar radiation and device operation). If $R_i, R_{ij}, C_i$ are not known from design specification, they can be obtained via model identification~\cite{JimMA08,BacM11}. In this paper, we consider the flow rate $m_i$ as the only control input to each zone, which is a common practice in others' work as well~\cite{HaoLK14,LinBMM15,HaoLKS15}.

\begin{prop}
When the HVAC system is off, i.e., $m_i=0$, system~(\ref{equ:thermalmodel}) asymptotically converges to an equilibrium point which is uniquely determined by the inputs $T^o, Q_i$. When the HVAC system is on, the asymptotic convergence property of system~(\ref{equ:thermalmodel}) remains and the equilibrium point is uniquely determined by the inputs $T^o, m_i, Q_i$.
\end{prop}

This proposition can be directly derived by rearranging~(\ref{equ:thermalmodel}) in state-space representation, and showing that the system matrix is Hurwitz. Therefore, the desiderata is to design $m_i$ only for periods when the HVAC system is on, more specifically, is to design the dynamics of $m_i$ to drive~(\ref{equ:thermalmodel}) to some desired state.

\subsection{The optimization problem}
In reality, each zone has a desired temperature which is the set point determined by users. The objective of the HVAC control considered in this paper is to regulate the temperature to be close to the set point in each zone, and to minimize the total energy consumption in the building. More specifically, we consider the following steady-state optimization problem:
\begin{subequations}\label{equ:opt}
\begin{gather}
\min_{Z_i,m_i} \sum_{i\in\mathcal{N}}\frac{1}{2}r_i(Z_i-T_i^{set})^2+\frac{w}{\eta}\sum_{i\in\mathcal{N}}c_am_i|Z_i-T^s| \nonumber\\
\qquad\qquad\qquad\qquad\quad +ws(\sum\nolimits_{i\in\mathcal{N}}m_i)^3 \\
\text{s. t. }\frac{T^o-Z_i}{R_i}+\sum_{j\in\mathcal{N}(i)}\frac{Z_j-Z_i}{R_{ij}}+c_am_i(T^s-Z_i)+Q_i=0 \\
T_i^{min}\le Z_i\le T_i^{max} \\
m_i^{min}\le m_i\le m_i^{max} \\
\sum\nolimits_{i\in\mathcal{N}}m_i\le\overline{m}
\end{gather}
\end{subequations}
where $i\in\mathcal{N}$ in~(\ref{equ:opt}b)-(\ref{equ:opt}d), $r_i, w$ are positive weight coefficients, $\eta$ is a given constant named Coefficient of Performance (COP, i.e., the ratio of the produced cold/heat to the consumed energy) of the cooling/heating coil, $s$ is a given coefficient regarding the fan power consumption, $T_i^{set}$ is the temperature set point satisfying $T_i^{min}<T_i^{set}<T_i^{max}$, $[T_i^{min},T_i^{max}]$ is the user comfort range, $[m_i^{min},m_i^{max}]$ is the range of $m_i$ in which $m_i^{min}$ is usually close to zero to guarantee a minimal ventilation level~\cite{MaAB11}, and $\overline{m}$ is the upper bound of the total flow rate in the building. Note that (i) to avoid confusion between steady-state values and temperature dynamics, we use $Z_i$ to denote the steady-state temperature value whereas $T_i$ is the temperature in the dynamic model~(\ref{equ:thermalmodel}), and (ii) $T^o$ and $Q_i$ are exogenous disturbances. We assume that problem~(\ref{equ:opt}) is feasible and satisfies Slater's condition~\cite{BoyV04}. Moreover, we have four important remarks for problem~(\ref{equ:opt}).

\noindent$\bullet$ In~(\ref{equ:opt}a), the term regarding the building energy consumption contains two parts, i.e., the energy consumption of the cooling/heating coil given by $\frac{1}{\eta}\sum_{i\in\mathcal{N}}c_am_i|Z_i-T^s|$ (weighted by $w$), and the power consumption of the supply fan which is approximately proportional to the cube of the total supply air flow~\cite{HaoLKS15}, i.e., $s(\sum\nolimits_{i\in\mathcal{N}}m_i)^3$ (weighted by $w$).

\noindent$\bullet$ Parameters $r_i, w$ are determined by users: if preferring more comfort, they can increase $r_i$ and decrease $w$, and vice versa.

\noindent$\bullet$ In the cooling mode, $T^s\ll Z_i (\text{or }T_i), \forall i$; in the heating mode, $T^s\gg Z_i (\text{or }T_i),\forall i$. This is usually true in practice. For example~\cite{HaoLKS15}, in the cooling mode, $T^s=12.8^{\circ}$C while $T_i\ge21.5^{\circ}$C. \emph{Note that once the mode is determined, the sign of $Z_i-T^s (\text{or }T_i-T^s)$ is determined.}

%T^s\ll T_i^{set}, T^s\gg T_i^{set}, T_i^{set},

\noindent$\bullet$ In~(\ref{equ:opt}e), we do not impose any lower bound constraint for the total flow rate as the lower bound usually equals $\sum_{i\in\mathcal{N}}m_i^{min}$. In addition, $\overline{m}<\sum_{i\in\mathcal{N}}m_i^{max}$ holds, otherwise, this constraint would become redundant.

To conclude, the goal is to design the regulating rule for $m_i$ in a \emph{decentralized} or \emph{distributed} way so that system~(\ref{equ:thermalmodel}) reaches a steady state which is an optimal solution to problem~(\ref{equ:opt}).

\begin{rem}
Though system~(\ref{equ:thermalmodel}) is a 1st-order RC model, using higher order RC models does not affect the formulation of~(\ref{equ:opt}) since it is a steady-state optimization problem. For example, for the 2nd-order model in~\cite{HaoLKS15,LinMB12}
\begin{gather}
C_i\dot T_i=\frac{T^o-T_i}{R_i}+\sum_{j\in\mathcal{N}(i)}\frac{T_{ij}-T_i}{R_{ij}}+c_am_i(T^s-T_i)+Q_i \nonumber\\
C_{ij}\dot T_{ij}=\frac{T_i-T_{ij}}{R_{ij}}+\frac{T_j-T_{ij}}{R_{ij}} \nonumber
\end{gather}
where $T_{ij}$ is the temperature of the wall separating zones $i$ and $j$, and $C_{ij}$ is the thermal capacitance of the wall, the corresponding steady-state Equation~(\ref{equ:opt}b) is given by
\begin{gather}
\frac{T^o-Z_i}{R_i}+\sum_{j\in\mathcal{N}(i)}\frac{Z_j-Z_i}{2R_{ij}}+c_am_i(T^s-Z_i)+Q_i=0 \nonumber
\end{gather}
which results from the steady-state equation $Z_{ij}=(Z_i+Z_j)/2$ ($Z_{ij}$ is the steady-state temperature of the wall), i.e., the steady-state equation of the higher order model can be reduced to formulation~(\ref{equ:opt}b) by eliminating states of solids in the building envelope. Because our control design procedures proposed later are based on solving the steady-state optimization problem~(\ref{equ:opt}), using higher order models will not affect them.
\end{rem}

%the thermal resistance of the wall separating zones, i.e., $R_{ij}$, is (sometimes much) larger than the thermal resistance of the wall\&window separating zones and outside, i.e., $R_i$ (actually we have assumed that each thermal zone has a window, so $R_{i}<R_{ij}$ holds).

\section{Method I: An Approximate Solution Procedure}\label{se:approximation}

\subsection{Approximation and convex relaxation}
Problem~(\ref{equ:opt}) is nonconvex because the quadratic equality constraint~(\ref{equ:opt}b) is nonconvex in $m_i, Z_i$. In reality, due to that $Z_i, Z_j$ (or $T_i, T_j$) of neighboring zones are often very close to each other and $R_{ij}$ is not small, the total heat gain from neighboring zones is (sometimes much) less dominant compared with the heat gain from the outside plus the indoor heat gain in every zone. Therefore, in this section, we ignore the term $\sum_{j\in\mathcal{N}(i)}\frac{T_j-T_i}{R_{ij}}$ in~(\ref{equ:thermalmodel}) as well as $\sum_{j\in\mathcal{N}(i)}\frac{Z_j-Z_i}{R_{ij}}$ in~(\ref{equ:opt}b). In addition, we approximate the power consumption of the supply fan by one of its upper bound $s\phi\sum_{i\in\mathcal{N}}\frac{1}{2}m_i^2$ where $\phi$ is a given constant satisfying $\phi\ge\overline{m}$, because using a Cauchy-Schwarz inequality, we have
\begin{align}
s\phi\sum_{i\in\mathcal{N}}\frac{1}{2}m_i^2\ge\frac{s\phi(\sum_{i\in\mathcal{N}}m_i)^2}{2|\mathcal{N}|}\ge\frac{s(\sum_{i\in\mathcal{N}}m_i)^3}{2|\mathcal{N}|} \nonumber
\end{align}
where $|\mathcal{N}|$ is the number of zones. By doing this, the objective function of the optimization problem becomes separable, which will enable us to design a decentralized controller later. To summarize, we obtain an approximate model to~(\ref{equ:thermalmodel}) as
\begin{align}\label{equ:thermalmodelapp}
C_i\dot T_i=\frac{T^o-T_i}{R_i}+c_am_i(T^s-T_i)+Q_i
\end{align}
together with an approximate problem to~(\ref{equ:opt}) given by
\begin{subequations}\label{equ:optapp}
\begin{gather}
\min_{Z_i,m_i} \sum_{i\in\mathcal{N}}\left[\frac{1}{2}r_i(Z_i-T_i^{set})^2+\frac{w}{\eta}c_am_i|Z_i-T^s|+\frac{1}{2}ws\phi m_i^2\right] \\
\text{s. t. }\frac{T^o-Z_i}{R_i}+c_am_i(T^s-Z_i)+Q_i=0 \\
T_i^{min}\le Z_i\le T_i^{max} \\
m_i^{min}\le m_i\le m_i^{max} \\
\sum\nolimits_{i\in\mathcal{N}}m_i\le\overline{m}
\end{gather}
\end{subequations}
where $i\in\mathcal{N}$ in~(\ref{equ:thermalmodelapp}),~(\ref{equ:optapp}b)-(\ref{equ:optapp}d). Remark that the approximate model~(\ref{equ:thermalmodelapp}) still keeps the global asymptomatic stability as in Proposition 1. We assume that problem~(\ref{equ:optapp}) is feasible and satisfies Slater's condition, moreover, \emph{we require $r_i>\frac{wc_a^2}{s\phi\eta^2}, i\in\mathcal{N}$ so that~(\ref{equ:optapp}a) can be strictly convex in $Z_i,m_i$}.

Although~(\ref{equ:optapp}) is still nonconvex due to the term $m_iZ_i$ in~(\ref{equ:optapp}b), we can actually convexify it and show that the convexification is indeed exact. From~(\ref{equ:optapp}b), we have $m_i=\frac{\frac{T^o-Z_i}{R_i}+Q_i}{c_a(Z_i-T^s)}$. Define
\begin{gather}
f_i(Z_i) \buildrel\Delta\over=\frac{\frac{T^o-Z_i}{R_i}+Q_i}{c_a(Z_i-T^s)}=m_i\ge m_i^{min}>0, \text{ }i\in\mathcal{N} \nonumber
\end{gather}
where the physical meaning of $f_i(Z_i)$ is the \emph{approximate} air flow rate for zone $i$ to stay at temperature $Z_i$. Then
\begin{gather}
f_i'(Z_i)=\frac{\frac{T^s-T^o}{R_i}-Q_i}{c_a(Z_i-T^s)^2}, \text{ }i\in\mathcal{N} \nonumber\\
f_i''(Z_i)=\frac{-2(\frac{T^s-T^o}{R_i}-Q_i)}{c_a(Z_i-T^s)^3}>0, \text{ }i\in\mathcal{N} \nonumber
\end{gather}
hold, i.e., $f_i(Z_i)$ is convex in $Z_i$, because $T^s\ll T^o$ holds in the cooling mode, and $T^s\gg T^o$, $\frac{T^s-T^o}{R_i}> Q_i$ hold (as $Q_i$ is less dominant~\cite{MaAB11,DenBM12}, especially in winter) in the heating mode. Therefore, we can relax~(\ref{equ:optapp}) as a convex optimization problem given by
\begin{subequations}\label{equ:optappre}
\begin{gather}
\min_{Z_i,m_i} \sum_{i\in\mathcal{N}}\left[\frac{1}{2}r_i(Z_i-T_i^{set})^2+\frac{w}{\eta}c_am_i|Z_i-T^s|+\frac{1}{2}ws\phi m_i^2\right] \\
\text{s. t. }\frac{\frac{T^o-Z_i}{R_i}+Q_i}{c_a(Z_i-T^s)}\le m_i, \text{ }i\in\mathcal{N} \\
T_i^{min}\le Z_i\le T_i^{max}, \text{ }i\in\mathcal{N} \\
m_i^{min}\le m_i\le m_i^{max}, \text{ }i\in\mathcal{N} \\
\sum\nolimits_{i\in\mathcal{N}}m_i\le\overline{m}.
\end{gather}
\end{subequations}

\begin{rem}\label{rem:1}
In the cooling mode, $f_i'(Z_i)<0, \forall i$ holds while in the heating mode, $f_i'(Z_i)>0, \forall i$ holds, i.e., $f_i(Z_i)$ is monotonic in $Z_i, \forall i$.
\end{rem}

\subsection{Tightness of the convex relaxation}
Before investigating the tightness of the convex relaxation, we make an assumption on the system parameters.

\begin{ass}\label{ass:1}
In the cooling mode, $T_i^{set}<T^o, \forall i$ holds; in the heating mode, $T_i^{set}>T^o, \forall i$ holds. Moreover, the zone temperature set point satisfies $f_i(T_i^{set})\ge m_i^{min}, \forall i$.
\end{ass}

This assumption holds in usual practice. Firstly, when the outdoor temperature is higher (in hot days)/lower (in cold days) than users' expectation (quantified as $T_i^{set}$), the HVAC system should be turned on. So we usually have $T_i^{set}<T^o, \forall i$ for cooling and $T_i^{set}>T^o, \forall i$ for heating. Secondly, note that $m_i^{min}\approx0, \forall i$ holds as stated earlier, and $f_i(Z_i)$ stands for the approximate flow rate for zone $i$ to stay at temperature $Z_i$. The inequality $f_i(T_i^{set})\ge m_i^{min}$ means that in order to make the temperature of zone $i$ be the set point, the flow rate needs to be no smaller than its minimum. Otherwise, we will have $f_i(T_i^{set})<m_i^{min}\approx0$ which can lead to $T_i^{set}\ge T^o$ in the cooling mode, and $T_i^{set}\le T^o$ in the heating mode, a contradiction to the first part of this assumption. To sum up, users can choose their set points to satisfy this assumption in reality, which is completely decentralized.

%(for example, in the cooling mode, if $T^o$ is close to users' expectation while indoor heat gains are large, they can choose $T_i^{set}$ lower than $T^o$ to satisfy this assumption)

The Lagrangian of problem~(\ref{equ:optappre}) is
\begin{align}
L=&\sum_{i\in\mathcal{N}}\left[\frac{1}{2}r_i(Z_i-T_i^{set})^2+\frac{w}{\eta}c_am_i|Z_i-T^s|+\frac{1}{2}ws\phi m_i^2\right] \nonumber\\
&+\sum_{i\in\mathcal{N}}\zeta_i\Bigg(\frac{\frac{T^o-Z_i}{R_i}+Q_i}{c_a(Z_i-T^s)}-m_i\Bigg)+\lambda^+\Big(\sum_{i\in\mathcal{N}}m_i-\overline m\Big) \nonumber\\
&+\sum_{i\in\mathcal{N}}\nu_i^+(Z_i-T_i^{max})+\sum_{i\in\mathcal{N}}\nu_i^-(T_i^{min}-Z_i) \nonumber\\
&+\sum_{i\in\mathcal{N}}\mu_i^+(m_i-m_i^{max})+\sum_{i\in\mathcal{N}}\mu_i^-(m_i^{min}-m_i) \nonumber
\end{align}
where $\zeta_i, \nu_i^+, \nu_i^-, \mu_i^+, \mu_i^-, \lambda^+$ are the Lagrange multipliers (dual variables) for constraints~(\ref{equ:optappre}b)-(\ref{equ:optappre}e). Since problem~(\ref{equ:optappre}) is convex, feasible and satisfies Slater's condition, the Karush-Kuhn-Tucker (KKT) conditions are necessary and sufficient conditions for optimality~\cite{BoyV04}, given by
\begin{subequations}\label{equ:kkt}
\begin{gather}
\frac{\partial L}{\partial Z_i}=r_i(Z_i-T_i^{set})\pm \frac{w}{\eta}c_am_i+\zeta_i\frac{\frac{T^s-T^o}{R_i}-Q_i}{c_a(Z_i-T^s)^2} \nonumber\\
\qquad\qquad\qquad\qquad\qquad\qquad\qquad +\nu_i^+-\nu_i^-=0 \\
\frac{\partial L}{\partial m_i}=ws\phi m_i\pm \frac{w}{\eta}c_a(Z_i-T^s)-\zeta_i+\mu_i^+-\mu_i^-+\lambda^+=0 \\
\zeta_i(f_i(Z_i)-m_i)=0, \text{ }\zeta_i\ge0, f_i(Z_i)-m_i\le0 \\
\nu_i^+(Z_i-T_i^{max})=0, \text{ }\nu_i^+\ge0, Z_i-T_i^{max}\le0 \\
\nu_i^-(T_i^{min}-Z_i)=0, \text{ }\nu_i^-\ge0, T_i^{min}-Z_i\le0 \\
\mu_i^+(m_i-m_i^{max})=0, \text{ }\mu_i^+\ge0, m_i-m_i^{max}\le0 \\
\mu_i^-(m_i^{min}-m_i)=0, \text{ }\mu_i^-\ge0, m_i^{min}-m_i\le0 \\
\lambda^+\Bigg(\sum_{i\in\mathcal{N}}m_i-\overline m\Bigg)=0, \text{ }\lambda^+\ge0, \sum_{i\in\mathcal{N}}m_i-\overline m\le0
\end{gather}
\end{subequations}
where $i\in\mathcal{N}$ in~(\ref{equ:kkt}a)-(\ref{equ:kkt}g), and ``$\pm$'' takes ``$+$'' in the cooling mode and ``$-$'' in the heating mode in~(\ref{equ:kkt}a)-(\ref{equ:kkt}b).

\begin{rem}\label{rem:2}
The convex relaxation from problem~(\ref{equ:optapp}) to~(\ref{equ:optappre}) is tight if and only if any solution of~(\ref{equ:kkt}) satisfies $\zeta_i>0$ or $\zeta_i=0, f_i(Z_i)=m_i$, $\forall i$.
\end{rem}

As for the tightness, we have the following theorem.

\begin{theo}\label{thm:1}
Under Assumption~\ref{ass:1}, the convex relaxation from problem~(\ref{equ:optapp}) to~(\ref{equ:optappre}) is always tight.
\end{theo}
$Proof$: Let us assume that there exists an $i$ in the solution of~(\ref{equ:kkt}) such that $\zeta_i=0, f_i(Z_i)<m_i$ holds. Then we can obtain $r_i(Z_i-T_i^{set})+\frac{w}{\eta}c_am_i=\nu_i^--\nu_i^+$ in the cooling mode and $r_i(Z_i-T_i^{set})-\frac{w}{\eta}c_am_i=\nu_i^--\nu_i^+$ in the heating mode from~(\ref{equ:kkt}a), and $\mu_i^-=ws\phi m_i\pm \frac{w}{\eta}c_a(Z_i-T^s)+\mu_i^++\lambda^+>0$ from~(\ref{equ:kkt}b). When $\nu_i^-=\nu_i^+=0$ holds, we have $Z_i<T_i^{set}$ in the cooling mode and $Z_i>T_i^{set}$ in the heating mode. When $\nu_i^->\nu_i^+=0$ holds which only happens in the cooling mode (otherwise in the heating mode, we end up with $Z_i=T_i^{min}>T_i^{set}$, a contradiction), we have $Z_i=T_i^{min}<T_i^{set}$. When $\nu_i^+>\nu_i^-=0$ holds which only happens in the heating mode, we have $Z_i=T_i^{max}>T_i^{set}$. These facts lead to $f_i(T_i^{set})<f_i(Z_i)<m_i=m_i^{min}$ by Remark~\ref{rem:1}, a contradiction to Assumption~\ref{ass:1}. Based on Remark~\ref{rem:2}, the convex relaxation from problem~(\ref{equ:optapp}) to~(\ref{equ:optappre}) is always tight. \qed

%\vspace{-0.02cm}
\subsection{A decentralized algorithm}
Theorem~\ref{thm:1} indicates that solving~(\ref{equ:optapp}) is equivalent to solving~(\ref{equ:optappre}), while~(\ref{equ:optappre}) can be solved in either a centralized or distributed/decentralized way. Any centralized algorithm requires to measure the outdoor temperature $T^o$ and the indoor heat gain $Q_i$ in every zone. Because these exogenous disturbances can fluctuate frequently and are not easy to obtain, the cost of centralized algorithms would be expensive. In this section, we develop a real-time decentralized algorithm that does not need measurement of these exogenous disturbances.

Since~(\ref{equ:optappre}) is convex, feasible and satisfies Slater's condition, we design the following algorithm to solve~(\ref{equ:optappre}) based on a standard primal-dual gradient method~\cite{FeiP10}:
\begin{subequations}\label{equ:controller}
\begin{align}
\dot Z_i=&-k_{Z_i}\Bigg(\frac{\partial L}{\partial Z_i}\Bigg)=k_{Z_i}\Bigg(r_i(T_i^{set}-Z_i)-\zeta_i\frac{\frac{T^s-T^o}{R_i}-Q_i}{c_a(Z_i-T^s)^2} \nonumber\\
&\mp \frac{w}{\eta}c_am_i-\nu_i^++\nu_i^-\Bigg) \\
\dot m_i=&-k_{m_i}\Bigg(\frac{\partial L}{\partial m_i}\Bigg)=k_{m_i}(-ws\phi m_i\mp \frac{w}{\eta}c_a(Z_i-T^s)+\zeta_i \nonumber\\
&-\mu_i^++\mu_i^--\lambda^+) \\
\dot\zeta_i=&k_{\zeta_i}\Bigg(\frac{\partial L}{\partial \zeta_i}\Bigg)_{\zeta_i}^+=k_{\zeta_i}\Bigg(\frac{\frac{T^o-Z_i}{R_i}+Q_i}{c_a(Z_i-T^s)}-m_i\Bigg)_{\zeta_i}^+ \\
\dot\nu_i^+=&k_{\nu_i^+}\Bigg(\frac{\partial L}{\partial \nu_i^+}\Bigg)_{\nu_i^+}^+=k_{\nu_i^+}(Z_i-T_i^{max})_{\nu_i^+}^+ \\
\dot\nu_i^-=&k_{\nu_i^-}\Bigg(\frac{\partial L}{\partial \nu_i^-}\Bigg)_{\nu_i^-}^+=k_{\nu_i^-}(T_i^{min}-Z_i)_{\nu_i^-}^+ \\
\dot\mu_i^+=&k_{\mu_i^+}\Bigg(\frac{\partial L}{\partial \mu_i^+}\Bigg)_{\mu_i^+}^+=k_{\mu_i^+}(m_i-m_i^{max})_{\mu_i^+}^+ \\
\dot\mu_i^-=&k_{\mu_i^-}\Bigg(\frac{\partial L}{\partial \mu_i^-}\Bigg)_{\mu_i^-}^+=k_{\mu_i^-}(m_i^{min}-m_i)_{\mu_i^-}^+ \\
\dot\lambda^+=&k_{\lambda^+}\Bigg(\frac{\partial L}{\partial \lambda^+}\Bigg)_{\lambda^+}^+=k_{\lambda^+}\Bigg(\sum_{i\in\mathcal{N}}m_i-\overline{m}\Bigg)_{\lambda^+}^+
\end{align}
\end{subequations}
where $i\in\mathcal{N}$ in~(\ref{equ:controller}a)-(\ref{equ:controller}g), $k_{Z_i}, k_{m_i}, k_{\zeta_i}, k_{\nu_i^+}, k_{\nu_i^-}, k_{\mu_i^+}, k_{\mu_i^-},$ $k_{\lambda^+}$ are positive scalars representing the controller gains, and ``$\mp$'' takes ``$-$'' in the cooling mode and ``$+$'' in the heating mode in~(\ref{equ:controller}a)-(\ref{equ:controller}b). Note that $T_i$ has its own dynamics given by~(\ref{equ:thermalmodelapp}) and thus can not be designed, which is why we replace $T_i$ with $Z_i$ initially, i.e., $Z_i, i\in\mathcal{N}$ are ancillary state variables. According to~\cite{FeiP10,CheMC16}, it is true that~(\ref{equ:controller}) asymptotically converges to an equilibrium point which is the unique optimal solution of~(\ref{equ:optappre}), since the optimization problem is convex and the objective function is strictly convex in the decision variables. Now using $m_i$ in~(\ref{equ:controller}b) as the input to system~(\ref{equ:thermalmodelapp}), we can naturally obtain a real-time decentralized controller to regulate~(\ref{equ:thermalmodelapp}) to a steady state which solves~(\ref{equ:optapp}).

\begin{theo}\label{thm:2}
Given constant/step change/slow-varying $T^o, Q_i$, under Assumption~\ref{ass:1}, the equilibrium point of the overall system~(\ref{equ:thermalmodelapp}) and~(\ref{equ:controller}) is unique, asymptotically stable and $T_i, m_i$ of the equilibrium point is the optimal solution of~(\ref{equ:optapp}).
\end{theo}
$Proof$: For constant/step change/slow-varying disturbances $T^o, Q_i$, since~(\ref{equ:optappre}) is convex and the objective function is strictly convex in $Z_i,m_i$, its optimal solution is unique~\cite{BoyV04}. Therefore, the equilibrium point of~(\ref{equ:controller}) is unique, asymptotically stable~\cite{FeiP10,CheMC16} and is the optimal solution of~(\ref{equ:optappre})/(\ref{equ:optapp}) by Theorem~\ref{thm:1}. So the equilibrium point of~(\ref{equ:thermalmodelapp}) and~(\ref{equ:controller}) is also asymptotically stable, due to the cascade nature, i.e.,~(\ref{equ:controller})$\rightarrow$(\ref{equ:thermalmodelapp}). Under the fact that the equilibrium point of~(\ref{equ:thermalmodelapp}) is uniquely determined by the inputs $T^o, m_i, Q_i$ where $m_i$ is given by~(\ref{equ:controller}b), we have $T_i=Z_i$ ($T_i$ is the state given by~(\ref{equ:thermalmodelapp})) when the overall system reaches steady state, which completes the proof. \qed

This theorem requires $T^o, Q_i$ to be either constant, step change, or slow-varying, which holds in practice as these disturbances vary at a time-scale of minutes. Remark that our controller operates in real-time, i.e., at a time-scale of seconds.

\begin{figure}[!t]
\centering
\includegraphics[width=0.41\textwidth]{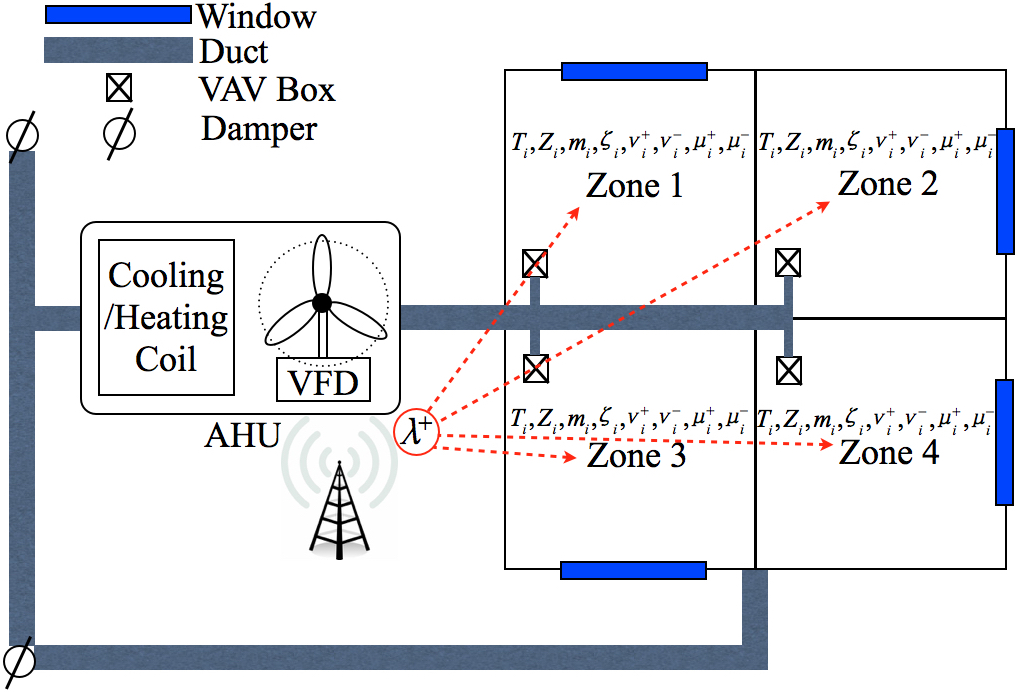}
\caption{Information exchange of the decentralized controller.}
\label{fig:2}
\end{figure}

\begin{figure*}[!t]
\centering
\begin{subfigure}[!h]{0.329\textwidth}
\centering
\includegraphics[width=1\textwidth]{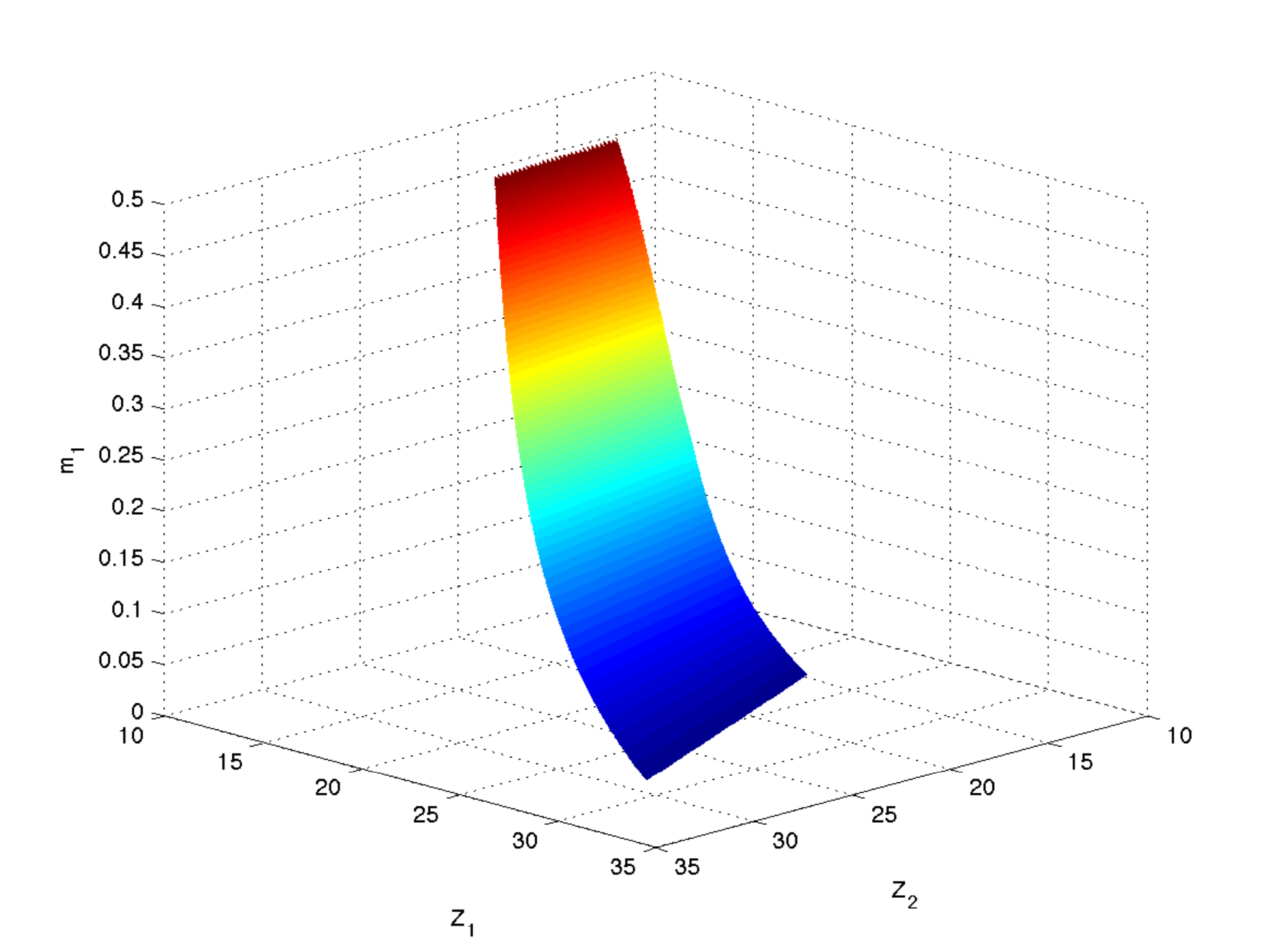}
\caption{}
\end{subfigure}
\begin{subfigure}[!h]{0.329\textwidth}
\centering
\includegraphics[width=1\textwidth]{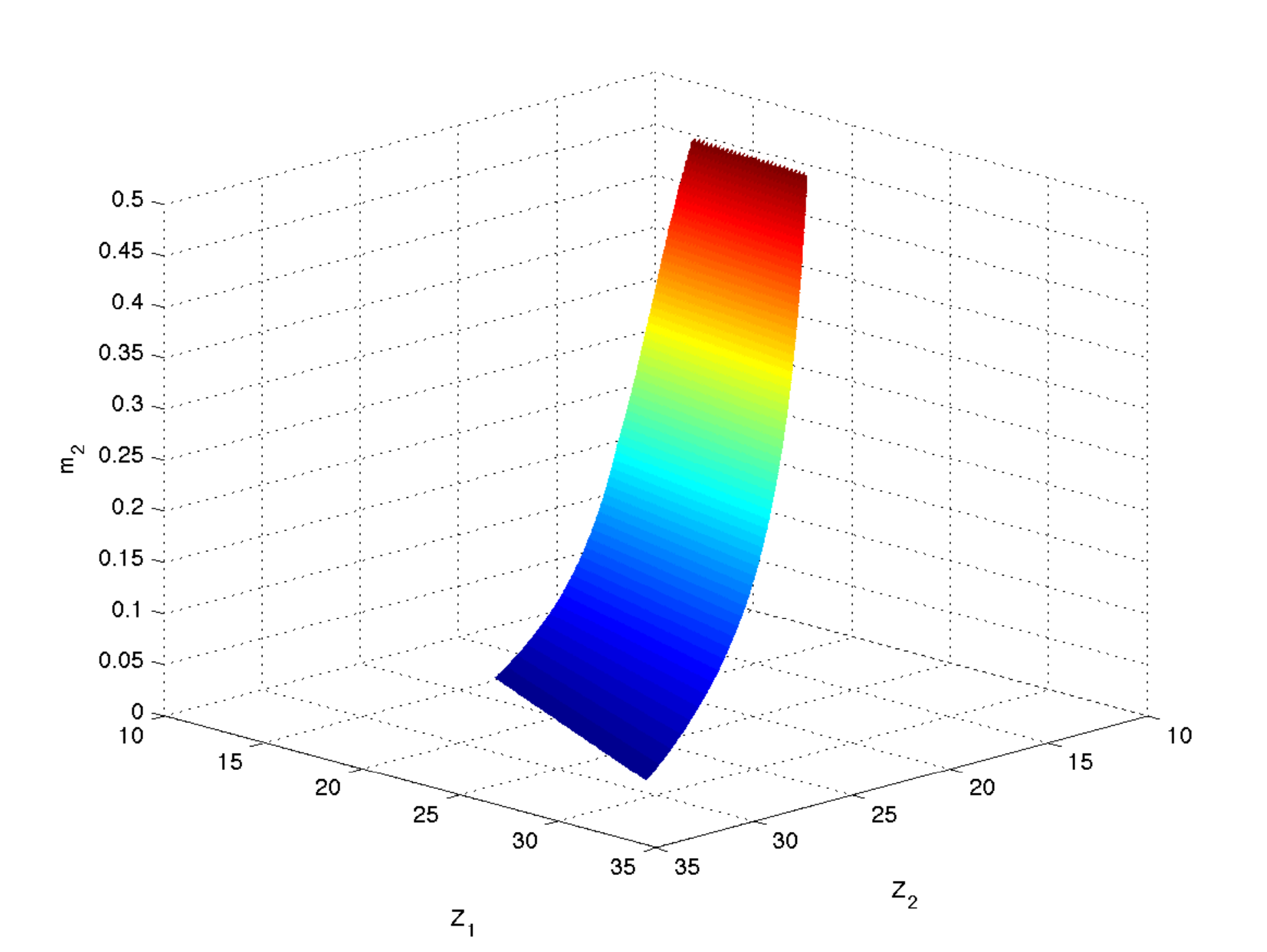}
\caption{}
\end{subfigure}
\begin{subfigure}[!h]{0.329\textwidth}
\centering
\includegraphics[width=1\textwidth]{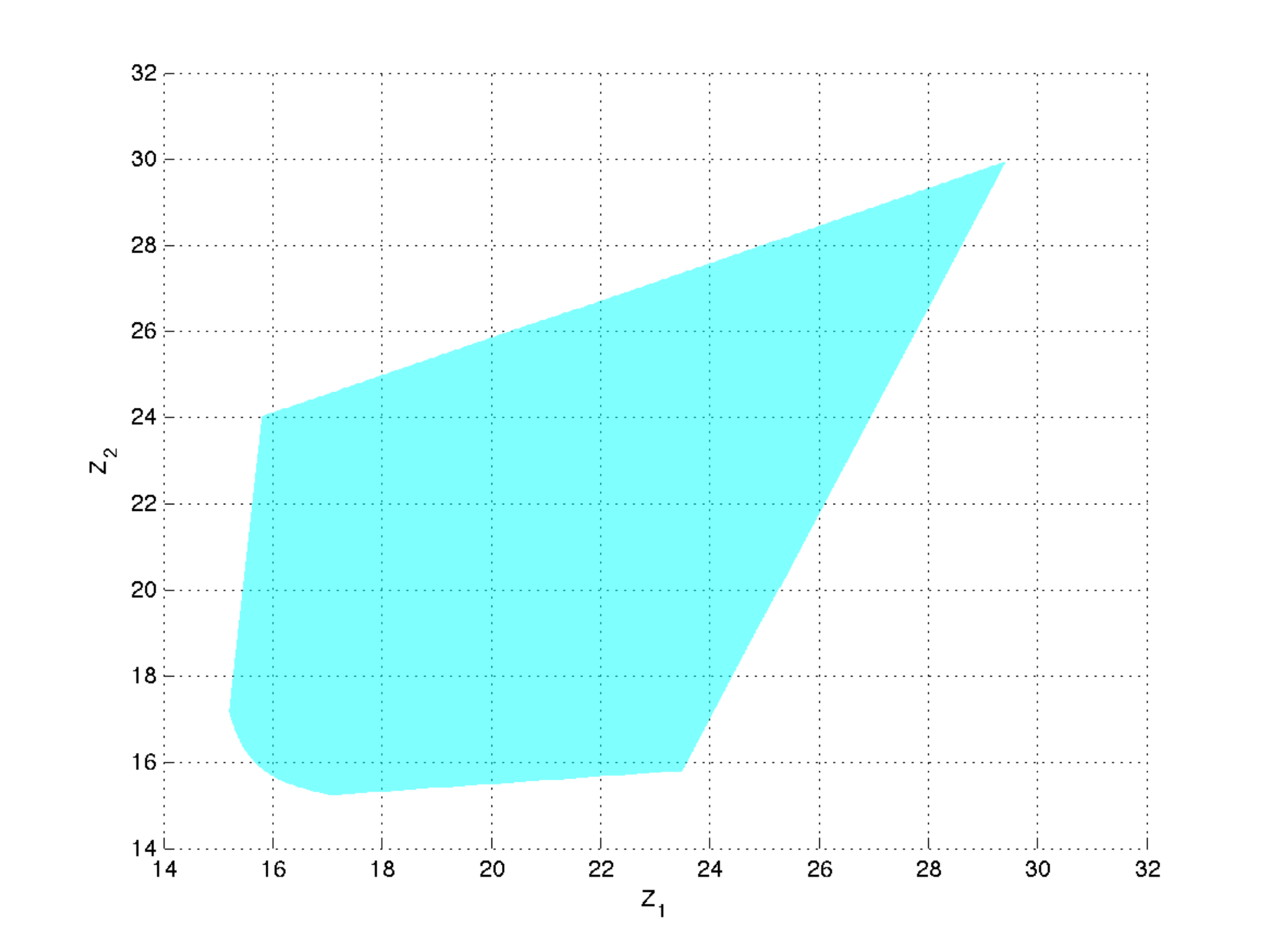}
\caption{}
\end{subfigure}
\caption{$m_1,m_2$ as functions of $Z_1,Z_2$. (a): $m_1$. (b): $m_2$. (c): the constraint set/domain for $Z_1,Z_2$.}
\label{fig:m1m2T1T2}
\end{figure*}

In~(\ref{equ:controller}a) and~(\ref{equ:controller}c), the disturbances $T^o, Q_i$ appear. Motivated by~\cite{ZhaLP15cdc}, to make the algorithm implementable without measuring these terms, we substitute~(\ref{equ:thermalmodelapp}) into~(\ref{equ:controller}a) and~(\ref{equ:controller}c) to get
\begin{subequations}\label{equ:controllerplus}
\begin{align}
\dot Z_i=&k_{Z_i}\Bigg(r_i(T_i^{set}-Z_i)\mp \frac{w}{\eta}c_am_i-\nu_i^++\nu_i^- \nonumber\\
&-\zeta_i\frac{\frac{T^s-T_i}{R_i}-C_i\dot T_i-c_am_i(T_i-T^s)}{c_a(Z_i-T^s)^2}\Bigg) \\
\dot\zeta_i=&k_{\zeta_i}\Bigg(\frac{\frac{T_i-Z_i}{R_i}+C_i\dot T_i+c_am_i(T_i-T^s)}{c_a(Z_i-T^s)}-m_i\Bigg)_{\zeta_i}^+
\end{align}
\end{subequations}
in which the derivative action $\dot T_i$ can be implemented by using a differentiator with some form of filtering~\cite{AngCL05}.

\textbf{Implementation.} The proposed control scheme~(\ref{equ:controller}b),~(\ref{equ:controller}d)-(\ref{equ:controller}h) and~(\ref{equ:controllerplus}) is completely decentralized as shown in Figure~\ref{fig:2} and can be implemented as follow. Given $T^s, R_i, C_i, r_i, w,$ $\phi, s, \eta, m_i^{min}, m_i^{max}$, each zone in the building collects $T_i^{set},$ $[T_i^{min},T_i^{max}]$ from users, locally measures its indoor temperature $T_i$, receives the feedback signal $\lambda^+$ from the supply fan/duct, and then uses these information to update $Z_i, m_i, \zeta_i,$ $\nu_i^+, \nu_i^-, \mu_i^+, \mu_i^-$ based on~(\ref{equ:controller}b),~(\ref{equ:controller}d)-(\ref{equ:controller}g) and~(\ref{equ:controllerplus}). On the other hand, given $\overline{m}$, the supply fan/duct locally measures the total flow rate $\sum_{i\in\mathcal{N}}m_i$ which is proportional to the fan speed~\cite{HaoLK14}, and uses~(\ref{equ:controller}h) to update $\lambda^+$, then broadcasts it. Here $T^s, R_i, C_i, s, \eta, m_i^{min}, m_i^{max}, \overline{m}$ are building parameters and $T_i^{set}, [T_i^{min},T_i^{max}], r_i, w, \phi$ are parameters specified by users.

\subsection{An extension}
In this section, we provide an alternative scenario which fits the above solution procedure. Consider a community or a neighborhood consisting of a number of separated houses. Each house is equipped with an HVAC system. Since these houses are separated (i.e., $R_{ij}=\infty$), we can naturally use system~(\ref{equ:thermalmodelapp}) to model their temperature dynamics. The objective is to regulate the temperature in each house to be close to its set point, meanwhile, to minimize the total energy consumption of the community. As a result, we end up with a steady-state optimization problem given by~(\ref{equ:optapp}). Note that (i) each \emph{entire} house is modeled by a RC model (we do not specify zones inside the house) and has a unique temperature set point determined by users; (ii) $m_i^{min}=0$ holds in~(\ref{equ:optapp}d) because we now allow houses to turn off their HVAC systems; (iii) $\overline{m}$ in~(\ref{equ:optapp}e) is an indicator of the upper bound of the total energy consumption in the community; (iv) the supply air temperature can be different in houses, i.e., $T^s$ can be replaced by $T_i^s$ in~(\ref{equ:thermalmodelapp}) and~(\ref{equ:optapp}a)-(\ref{equ:optapp}b) (the mixed mode case, i.e., some of HVAC systems are in the cooling mode while the others are in the heating mode, is naturally included), and similar for $w,\eta,s,\phi$.

Correspondingly, we modify Assumption~\ref{ass:1} a bit as follow.

\begin{ass}\label{ass:2}
For any house $i$, when the HVAC system is on, the house temperature set point satisfies $f_i(T_i^{set})>0, \forall i$. Otherwise, the HVAC in house $i$ is turned off and all terms and constraints relating to house $i$ in problem~(\ref{equ:optapp})/(\ref{equ:optappre}) are excluded.
\end{ass}

As before, users can choose their set points to satisfy this assumption, which is decentralized. Under this assumption and following the proof of Theorem~\ref{thm:1}, we can still ensure the tightness of the convex relaxation procedure in Section~\ref{se:approximation}.A. Moreover, we can still derive a decentralized algorithm similar to~(\ref{equ:controller})-(\ref{equ:controllerplus}) for HVAC system management in the community. What needs to be emphasized is that now~(\ref{equ:controller}h) is updated in a control center of the community, which receives $m_i$ from each house and sends $\lambda^+$ to them (when $m_i=0$, the control center can know that house $i$ has turned off its HVAC system).

\section{Method II: The General Case}\label{se:general}
In this section, we focus on solving problem~(\ref{equ:opt}) directly. The key step is to reformulate~(\ref{equ:opt}) as an optimization problem without using the decision variable $m_i$ by substituting~(\ref{equ:opt}b) into~(\ref{equ:opt}a) and~(\ref{equ:opt}d)-(\ref{equ:opt}e) to eliminate $m_i$. Before showing the details, let us first consider a simple scenario, i.e., a building with two adjacent zones, to gain an insight into this method.

\subsection{An illustrating example}

\begin{figure}[!t]
\centering
\includegraphics[width=0.41\textwidth]{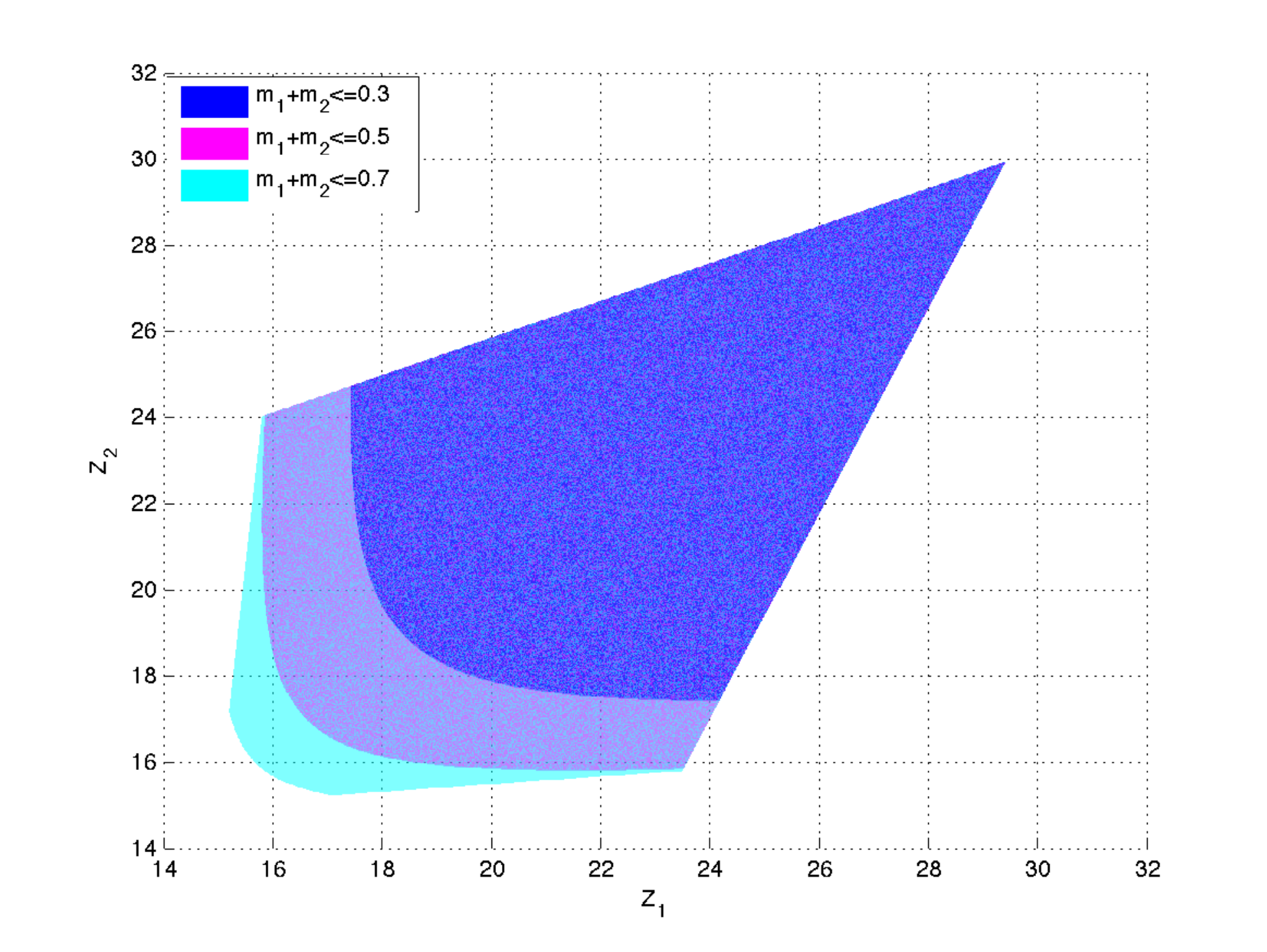}
\caption{The constraint sets for $Z_1,Z_2$ when the upper bound of $m_1+m_2$ changes.}
\label{fig:T1T2rrr}
\end{figure}

Let the constraint set of problem~(\ref{equ:opt}) be
\begin{gather}
\frac{30-Z_1}{15}+\frac{Z_2-Z_1}{18}+1.012m_1(12.8-Z_1)+0.1=0 \nonumber\\
\frac{30-Z_2}{16}+\frac{Z_1-Z_2}{18}+1.012m_2(12.8-Z_2)+0.2=0 \nonumber\\
0.01\le m_1\le0.5,\quad
0.01\le m_2\le0.5,\quad
m_1+m_2\le0.7. \nonumber
\end{gather}
where we do not add user comfort constraint~(\ref{equ:opt}c) (the reason will become clear later). The domain for $m_1,m_2, Z_1,Z_2$ is shown in Figure~\ref{fig:m1m2T1T2}(a)-(b), which are nonconvex surfaces. However, by rewriting $m_1,m_2$ as functions of $Z_1,Z_2$
\begin{gather}
m_1=\frac{\frac{30-Z_1}{15}+\frac{Z_2-Z_1}{18}+0.1}{1.012(Z_1-12.8)},\quad
m_2=\frac{\frac{30-Z_2}{16}+\frac{Z_1-Z_2}{18}+0.2}{1.012(Z_2-12.8)} \nonumber
\end{gather}
and substituting them into all three inequality constraints, we can see that the constraint set for $Z_1,Z_2$ is actually convex as shown in Figure~\ref{fig:m1m2T1T2}(c), which is also the two-dimensional view of Figure~\ref{fig:m1m2T1T2}(a)-(b). The reason is as follows. Since the cooling mode is considered (i.e., $Z_1, Z_2\gg T^s=12.8$), the constraints $0.01\le m_1\le0.5$ and $0.01\le m_2\le0.5$ are equivalent to
\begin{align}
0.01\times1.012(Z_1-12.8)\le&\frac{30-Z_1}{15}+\frac{Z_2-Z_1}{18}+0.1 \nonumber\\
\le&0.5\times1.012(Z_1-12.8) \nonumber\\
0.01\times1.012(Z_2-12.8)\le&\frac{30-Z_2}{16}+\frac{Z_1-Z_2}{18}+0.2 \nonumber\\
\le&0.5\times1.012(Z_2-12.8) \nonumber
\end{align}
which are linear, and therefore, convex. So the convexity of the constraint set only depends on the convexity of the function $m_1+m_2$. Then we calculate the Hessian matrix of
\begin{gather}
m_1+m_2=\frac{\frac{30-Z_1}{15}+\frac{Z_2-Z_1}{18}+0.1}{1.012(Z_1-12.8)} +\frac{\frac{30-Z_2}{16}+\frac{Z_1-Z_2}{18}+0.2}{1.012(Z_2-12.8)} \nonumber
\end{gather}
with respect to $Z_1,Z_2$. Any point in the domain in Figure~\ref{fig:m1m2T1T2}(c) has a positive semi-definite Hessian matrix of $m_1+m_2$.
%satisfies that the Hessian matrix at that point is positive semi-definite, which ensures the convexity.

Next, we modify the upper bound of $m_1+m_2$ to be $0.5$ and $0.3$, and draw the constraint sets of $Z_1,Z_2$, as illustrated in Figure~\ref{fig:T1T2rrr}. It can be seen that these constraint sets are also convex, because any point in the domains displayed in Figure~\ref{fig:T1T2rrr} has a positive semi-definite Hessian matrix of $m_1+m_2$. Note that including user comfort constraint~(\ref{equ:opt}c) will not affect the above analysis since it does not affect the convexity of the constraint set of $Z_1,Z_2$ but only makes the set smaller, i.e., the above analysis is actually less conservative than that with constraint~(\ref{equ:opt}c). These numerical results indicate that once we exclude $m_i$ from problem~(\ref{equ:opt}), convexity can be obtained.

\subsection{Solution procedure}
Now we provide the solution procedure for~(\ref{equ:opt}). Define
\begin{gather}
h(Z)\buildrel\Delta\over=\sum_{i\in\mathcal{N}}m_i=\sum_{i\in\mathcal{N}}\frac{\frac{T^o-Z_i}{R_i}+\sum_{j\in\mathcal{N}(i)}\frac{Z_j-Z_i}{R_{ij}}+Q_i}{c_a(Z_i-T^s)}>0. \nonumber
\end{gather}
where $Z$ is a collection of $Z_i, i\in\mathcal{N}$. By substituting~(\ref{equ:opt}b) into~(\ref{equ:opt}a) and~(\ref{equ:opt}d)-(\ref{equ:opt}e) to eliminate $m_i$, problem~(\ref{equ:opt}) becomes
\begin{subequations}\label{equ:opts=0ZZZ}
\begin{gather}
\min_{Z_i}\sum_{i\in\mathcal{N}}\frac{1}{2}r_i(Z_i-T_i^{set})^2+\frac{w}{\eta}\left|\sum_{i\in\mathcal{N}}\frac{T^o-Z_i}{R_i}+Q_i\right| \nonumber\\
\qquad\qquad\qquad\qquad\quad +ws(h(Z))^3 \\
\text{s. t. } T_i^{min}\le Z_i\le T_i^{max} \\
m_i^{min}\le \frac{\frac{T^o-Z_i}{R_i}+\sum_{j\in\mathcal{N}(i)}\frac{Z_j-Z_i}{R_{ij}}+Q_i}{c_a(Z_i-T^s)}\le m_i^{max} \\
h(Z)\le\overline{m}.
\end{gather}
\end{subequations}
where $i\in\mathcal{N}$ in~(\ref{equ:opts=0ZZZ}b)-(\ref{equ:opts=0ZZZ}c). Constraint~(\ref{equ:opts=0ZZZ}c) can actually become linear once the cooling/heating mode (the sign of $Z_i-T^s$) is determined (also, the absolute value sign in~(\ref{equ:opts=0ZZZ}a) can be removed). The convexity of the constraint set only depends on constraint~(\ref{equ:opts=0ZZZ}d). Now we claim that the HVAC system satisfies the following assumption under normal operating conditions.

\begin{ass}\label{ass:3}
The Hessian matrix of $h(Z)$, i.e.,
\begin{gather}
\frac{\partial^2 h(Z)}{\partial Z^2}=\left[ {\begin{array}{*{20}{c}}
{\frac{{{\partial ^2}h(Z)}}{{\partial Z_i^2}}}&{\frac{{{\partial ^2}h(Z)}}{{\partial {Z_i}\partial {Z_j}}}}\\
{\frac{{{\partial ^2}h(Z)}}{{\partial {Z_j}\partial {Z_i}}}}&{\frac{{{\partial ^2}h(Z)}}{{\partial Z_j^2}}}
\end{array}} \right] \nonumber\\
\frac{\partial ^2 h(Z)}{\partial Z_i^2}=\frac{-2(\frac{T^s-T^o}{R_i}+\sum_{j\in\mathcal{N}(i)}\frac{T^s-Z_j}{R_{ij}}-Q_i)}{c_a(Z_i-T^s)^3}>0 \nonumber\\
\frac{\partial ^2 h(Z)}{\partial Z_i\partial Z_j}=-\frac{1}{R_{ij}c_a(Z_i-T^s)^2}-\frac{1}{R_{ij}c_a(Z_j-T^s)^2}<0 \nonumber
\end{gather}
is positive semi-definite.
\end{ass}

Assumption~\ref{ass:3} is not conservative and it usually holds in practice, because $\frac{\partial^2 h(Z)}{\partial Z^2}$ is usually diagonal dominant due to that $Z_i,Z_j$ of neighboring zones do not deviate too much from each other (this can also be guaranteed by properly adjusting user comfort range), i.e.,
\begin{gather}
\left|\frac{\partial ^2 h(Z)}{\partial Z_i^2}\right|-\sum_{j\in\mathcal{N}(i)}\left|\frac{\partial ^2 h(Z)}{\partial Z_i\partial Z_j}\right|\approx\frac{-2(\frac{T^s-T^o}{R_i}-Q_i)}{c_a(Z_i-T^s)^3}>0. \nonumber
\end{gather}
This fact is demonstrated in the previous example: even though $Z_1,Z_2$ deviate about $8^{\circ}$C from each other, the constraint set remains convex as shown in Figures~\ref{fig:m1m2T1T2}(c)-\ref{fig:T1T2rrr}.

Under Assumption~\ref{ass:3} (and the sign of $Z_i-T^s$ is determined), problem~(\ref{equ:opts=0ZZZ}) is convex since the objective function~(\ref{equ:opts=0ZZZ}a) is strictly convex in $Z_i$ as convexity is preserved from $h(Z)>0$ to $(h(Z))^3$~\cite{BoyV04}. Motivated by a standard primal-dual gradient method~\cite{FeiP10}, we design the following algorithm to solve~(\ref{equ:opts=0ZZZ}) (assuming the cooling mode here, i.e., $Z_i>T^s$, the heating mode case is similar):
\begin{subequations}\label{equ:controls=0}
\begin{align}
\dot Z_i=&k_{Z_i}\Bigg(r_i(T_i^{set}-Z_i)+\mu_i^+\Big(\frac{1}{R_i}+\sum_{j\in\mathcal{N}(i)}\frac{1}{R_{ij}}+m_i^{max}c_a\Big) \nonumber\\
&-\sum_{j\in\mathcal{N}(i)}\frac{\mu_j^+}{R_{ij}}-\mu_i^-\Big(\frac{1}{R_i}+\sum_{j\in\mathcal{N}(i)}\frac{1}{R_{ij}}+m_i^{min}c_a\Big)-\nu_i^+ \nonumber\\
&+\sum_{j\in\mathcal{N}(i)}\frac{\mu_j^-}{R_{ij}}+\nu_i^--\Bigg(\frac{\frac{T^s-T^o}{R_i}+\sum_{j\in\mathcal{N}(i)}\frac{T^s-Z_j}{R_{ij}}-Q_i}{c_a(Z_i-T^s)^2} \nonumber\\
&+\sum_{j\in\mathcal{N}(i)}\frac{1}{R_{ij}c_a(Z_j-T^s)}\Bigg)(3wsh^2(Z)+\lambda^+)+\frac{w}{\eta R_i}\Bigg) \\
\dot\nu_i^+=&k_{\nu_i^+}(Z_i-T_i^{max})_{\nu_i^+}^+ \\
\dot\nu_i^-=&k_{\nu_i^-}(T_i^{min}-Z_i)_{\nu_i^-}^+ \\
\dot\mu_i^+=&k_{\mu_i^+}\Bigg(\frac{T^o-Z_i}{R_i}+\sum_{j\in\mathcal{N}(i)}\frac{Z_j-Z_i}{R_{ij}}+Q_i \nonumber\\
&-m_i^{max}c_a(Z_i-T^s)\Bigg)_{\mu_i^+}^+ \\
\dot\mu_i^-=&k_{\mu_i^-}\Bigg(m_i^{min}c_a(Z_i-T^s)-\frac{T^o-Z_i}{R_i} \nonumber\\
&-\sum_{j\in\mathcal{N}(i)}\frac{Z_j-Z_i}{R_{ij}}-Q_i\Bigg)_{\mu_i^-}^+ \\
\dot\lambda^+=&k_{\lambda^+}\Bigg(\sum_{i\in\mathcal{N}}\frac{\frac{T^o-Z_i}{R_i}+\sum_{j\in\mathcal{N}(i)}\frac{Z_j-Z_i}{R_{ij}}+Q_i}{c_a(Z_i-T^s)}-\overline{m}\Bigg)_{\lambda^+}^+
\end{align}
\end{subequations}
where $\nu_i^+, \nu_i^-, \mu_i^+, \mu_i^-, \lambda^+$ are the Lagrange multipliers (dual variables) for constraints~(\ref{equ:opts=0ZZZ}b)-(\ref{equ:opts=0ZZZ}d), $k_{Z_i}, k_{\nu_i^+}, k_{\nu_i^-}, k_{\mu_i^+}, k_{\mu_i^-},$ $k_{\lambda^+}$ are positive scalars representing the controller gains, and $i\in\mathcal{N}$ in~(\ref{equ:controls=0}a)-(\ref{equ:controls=0}e). In addition, we adopt the following low pass dynamics for each $m_i$ as the control input to system~(\ref{equ:thermalmodel}):
\begin{gather}\label{equ:controls=0mi}
\dot m_i=k_{m}\Bigg(\frac{\frac{T^o-Z_i}{R_i}+\sum_{j\in\mathcal{N}(i)}\frac{Z_j-Z_i}{R_{ij}}+Q_i}{c_a(Z_i-T^s)}-m_i\Bigg)
\end{gather}
where $k_{m}>0$ is the homogeneous controller gain. The reason for using the low pass dynamics is that it can attenuate high frequency noises to help improve system performance.

\begin{theo}
Given constant/step change/slow-varying $T^o, Q_i$ (remark that they vary at a time-scale of minutes), under Assumption~\ref{ass:3}, any trajectory of system~(\ref{equ:thermalmodel}),~(\ref{equ:controls=0})-(\ref{equ:controls=0mi}) asymptotically converges to a unique equilibrium point at which problem~(\ref{equ:opts=0ZZZ})/(\ref{equ:opt}) is solved and $T_i=Z_i, \forall i$ (this is because of the strict convexity of the function~(\ref{equ:opts=0ZZZ}a) in $Z_i$ and the cascade nature of the overall system, i.e.,~(\ref{equ:controls=0})$\rightarrow$(\ref{equ:controls=0mi})$\rightarrow$(\ref{equ:thermalmodel})).
\end{theo}

In~(\ref{equ:controls=0})-(\ref{equ:controls=0mi}), the disturbances $T^o, Q_i$ appear. Motivated by~\cite{ZhaLP15cdc}, to make the algorithm implementable without measuring these terms, we substitute~(\ref{equ:thermalmodel}) into~(\ref{equ:controls=0})-(\ref{equ:controls=0mi}) to get
\begin{subequations}\label{equ:controls=0plus}
\begin{align}
&\dot Z_i=k_{Z_i}\Bigg(r_i(T_i^{set}-Z_i)+\frac{w}{\eta R_i}-\nu_i^++\nu_i^- \nonumber\\
&+\mu_i^+\Big(\frac{1}{R_i}+\sum_{j\in\mathcal{N}(i)}\frac{1}{R_{ij}}+m_i^{max}c_a\Big)-\sum_{j\in\mathcal{N}(i)}\frac{\mu_j^+}{R_{ij}} \nonumber\\
&-\mu_i^-\Big(\frac{1}{R_i}+\sum_{j\in\mathcal{N}(i)}\frac{1}{R_{ij}}+m_i^{min}c_a\Big)+\sum_{j\in\mathcal{N}(i)}\frac{\mu_j^-}{R_{ij}} \nonumber\\
&-(3wsh^2(Z)+\lambda^+)\Bigg(\sum_{j\in\mathcal{N}(i)}\frac{1}{R_{ij}c_a(Z_j-T^s)})+ \nonumber\\
&\frac{\frac{T^s-T_i}{R_i}-C_i\dot T_i+\sum\limits_{j\in\mathcal{N}(i)}\frac{T^s-Z_j+T_j-T_i}{R_{ij}}-c_am_i(T_i-T^s)}{c_a(Z_i-T^s)^2}\Bigg)\Bigg) \\
&\dot\mu_i^+=k_{\mu_i^+}\Bigg(\frac{T_i-Z_i}{R_i}+C_i\dot T_i+\sum_{j\in\mathcal{N}(i)}\frac{Z_j-Z_i-T_j+T_i}{R_{ij}} \nonumber\\
&+c_am_i(T_i-T^s)-m_i^{max}c_a(Z_i-T^s)\Bigg)_{\mu_i^+}^+ \\
&\dot\mu_i^-=k_{\mu_i^-}\Bigg(m_i^{min}c_a(Z_i-T^s)-\frac{T_i-Z_i}{R_i}-C_i\dot T_i \nonumber\\
&-\sum_{j\in\mathcal{N}(i)}\frac{Z_j-Z_i-T_j+T_i}{R_{ij}}-c_am_i(T_i-T^s)\Bigg)_{\mu_i^-}^+ \\
&\dot\lambda^+=k_{\lambda^+}(h(Z)-\overline{m})_{\lambda^+}^+ \\
&\dot m_i=k_{m}\Bigg(-m_i+ \nonumber\\
&\frac{\frac{T_i-Z_i}{R_i}+C_i\dot T_i+\sum\limits_{j\in\mathcal{N}(i)}\frac{Z_j-Z_i-T_j+T_i}{R_{ij}}+c_am_i(T_i-T^s)}{c_a(Z_i-T^s)}\Bigg) \\
&h(Z)=\sum_{i\in\mathcal{N}} \nonumber\\
&\frac{\frac{T_i-Z_i}{R_i}+C_i\dot T_i+\sum\limits_{j\in\mathcal{N}(i)}\frac{Z_j-Z_i-T_j+T_i}{R_{ij}}+c_am_i(T_i-T^s)}{c_a(Z_i-T^s)}
\end{align}
\end{subequations}
in which the derivative action $\dot T_i$ can be implemented by using a differentiator with some form of filtering~\cite{AngCL05}.

\begin{figure}[!t]
\centering
\includegraphics[width=0.43\textwidth]{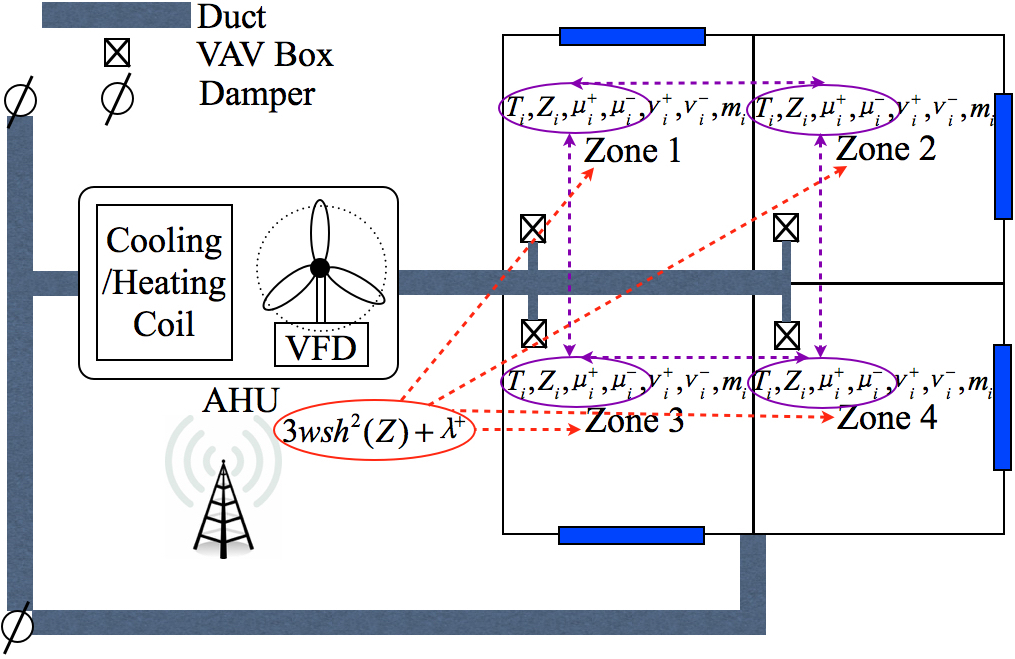}
\caption{Information exchange of the distributed controller.}
\label{fig:3}
\end{figure}

\textbf{Implementation.} The proposed controller~(\ref{equ:controls=0}b)-(\ref{equ:controls=0}c) and~(\ref{equ:controls=0plus}) is completely distributed as shown in Figure~\ref{fig:3} and can be implemented as follow. Given $T^s, R_i, R_{ij},C_i, r_i, w, s, \eta,$ $m_i^{min}, m_i^{max}$, each zone in the building collects $T_i^{set},$ $[T_i^{min},T_i^{max}]$ from users, locally measures its indoor temperature $T_i$, receives the feedback signals $3wsh^2(Z)+\lambda^+$ from the supply fan/duct and $T_j, Z_j, \mu_j^+, \mu_j^-$ from its neighboring zones, and then uses these information to update $Z_i, \nu_i^+, \nu_i^-, \mu_i^+, \mu_i^-, m_i$ based on~(\ref{equ:controls=0}b)-(\ref{equ:controls=0}c),~(\ref{equ:controls=0plus}a)-(\ref{equ:controls=0plus}c) and~(\ref{equ:controls=0plus}e). On the other hand, given $\overline{m}$, the supply fan/duct locally measures the total flow rate $\sum_{i\in\mathcal{N}}m_i$ which is proportional to the fan speed~\cite{HaoLK14}, and uses~(\ref{equ:controls=0plus}d) to update $\lambda^+$, and then broadcasts $3wsh^2(Z)+\lambda^+$. Note that the supply fan/duct can compute $\lambda^+$ and $h(Z)$ via
\begin{subequations}
\begin{gather}
\dot\lambda^+=k_{\lambda^+}(h(Z)-\overline{m})_{\lambda^+}^+ \\
h(Z)=\frac{1}{k_{m}}\sum_{i\in\mathcal{N}}\dot m_i+\sum_{i\in\mathcal{N}}m_i.
\end{gather}
\end{subequations}
Also, $T^s, R_i, R_{ij}, C_i, s, \eta, m_i^{min}, m_i^{max}, \overline{m}$ are building parameters and $T_i^{set}, [T_i^{min},T_i^{max}], r_i, w$ are given by users.

\section{Numerical Investigations}\label{se:simulation}
In this section, we present two numerical examples for scenarios described in Sections~\ref{se:approximation}-\ref{se:general} respectively, using a house with four adjacent zones as illustrated in Figure~\ref{fig:1}. Only the cooling case is presented in the following, while the heating case is similar under the proposed control schemes.

\begin{figure}[!t]
\centering
\includegraphics[width=0.5\textwidth]{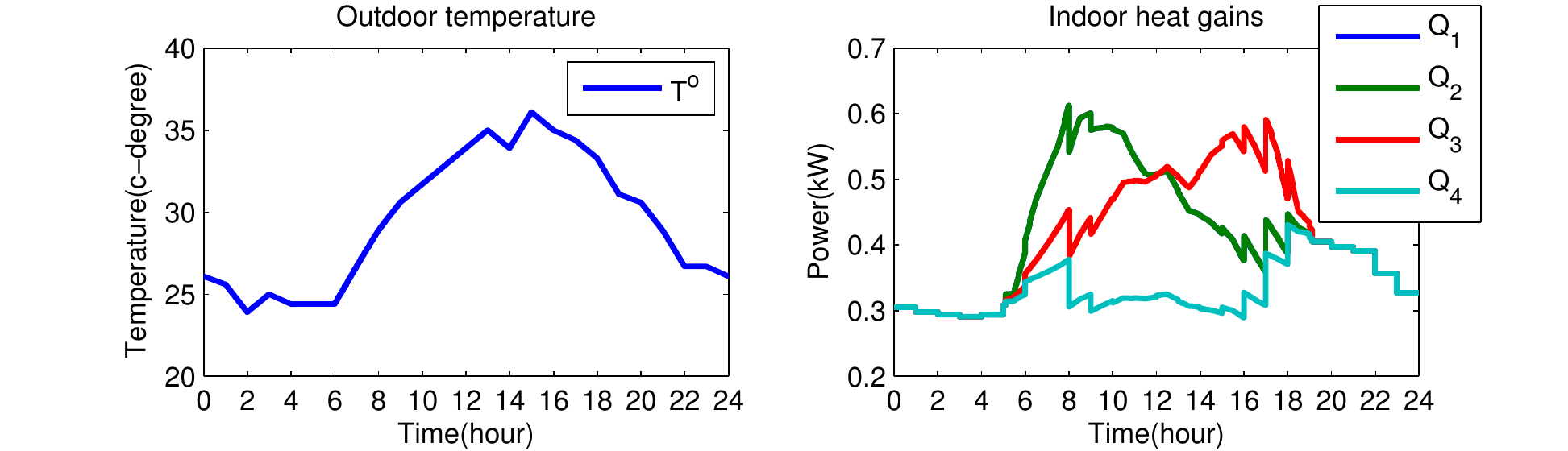}
\caption{Profiles of the disturbances ($Q_1=Q_2$).}
\label{fig:disturb}
\end{figure}

The parameters of the simulations are from~\cite{DerFS11,LiWB16}: all $C_i=20$kJ$\text{/}^{\circ}$C, all $R_i=15^{\circ}$C/kW, all $R_{ij}=23^{\circ}$C/kW, $c_a=1.012$kJ/kg$\text{/}^{\circ}$C, $T^s=12.8^{\circ}$C (cooling), $s=2$kW/$\text{(kg/s)}^3$, $\eta=2.9$, all comfort ranges are within $\pm 1.5^{\circ}$C of their set points, all $[m_i^{min},m_i^{max}]=[0.01,0.45]$kg/s, $\overline{m}=0.5$kg/s, $\phi=1$kg/s, all $r_i=0.1$p.u., $w=1$p.u., all $k_{Z_i}=0.067$p.u., all $k_{m_i}=k_{\zeta_i}=k_{\nu_i^+}=k_{\nu_i^-}=k_{\mu_i^+}=k_{\mu_i^-}=k_{\lambda^+}=k_{m}=1$p.u. (i.e., per unit), and the disturbance injection is shown in Figure~\ref{fig:disturb}.

\begin{figure}[!t]
\centering
\includegraphics[width=0.5\textwidth]{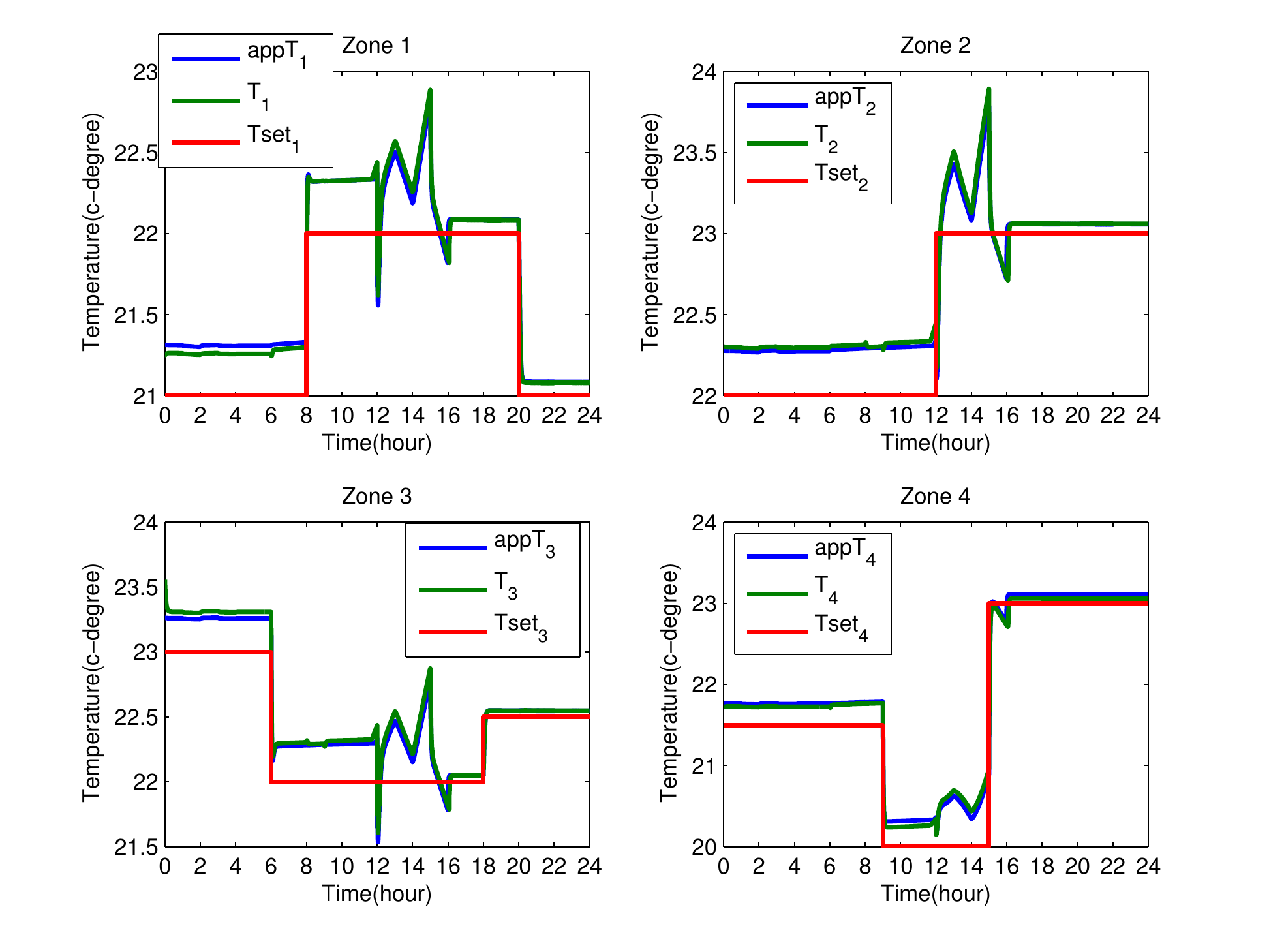}
\caption{Temperatures in each zone under controller~(\ref{equ:controller}b),~(\ref{equ:controller}d)-(\ref{equ:controller}h) and~(\ref{equ:controllerplus}): curves labeled with ``app'' indicate the case of using the approximate model~(\ref{equ:thermalmodelapp}).}
\label{fig:tempapp}
\end{figure}

\begin{figure}[!t]
\centering
\includegraphics[width=0.5\textwidth]{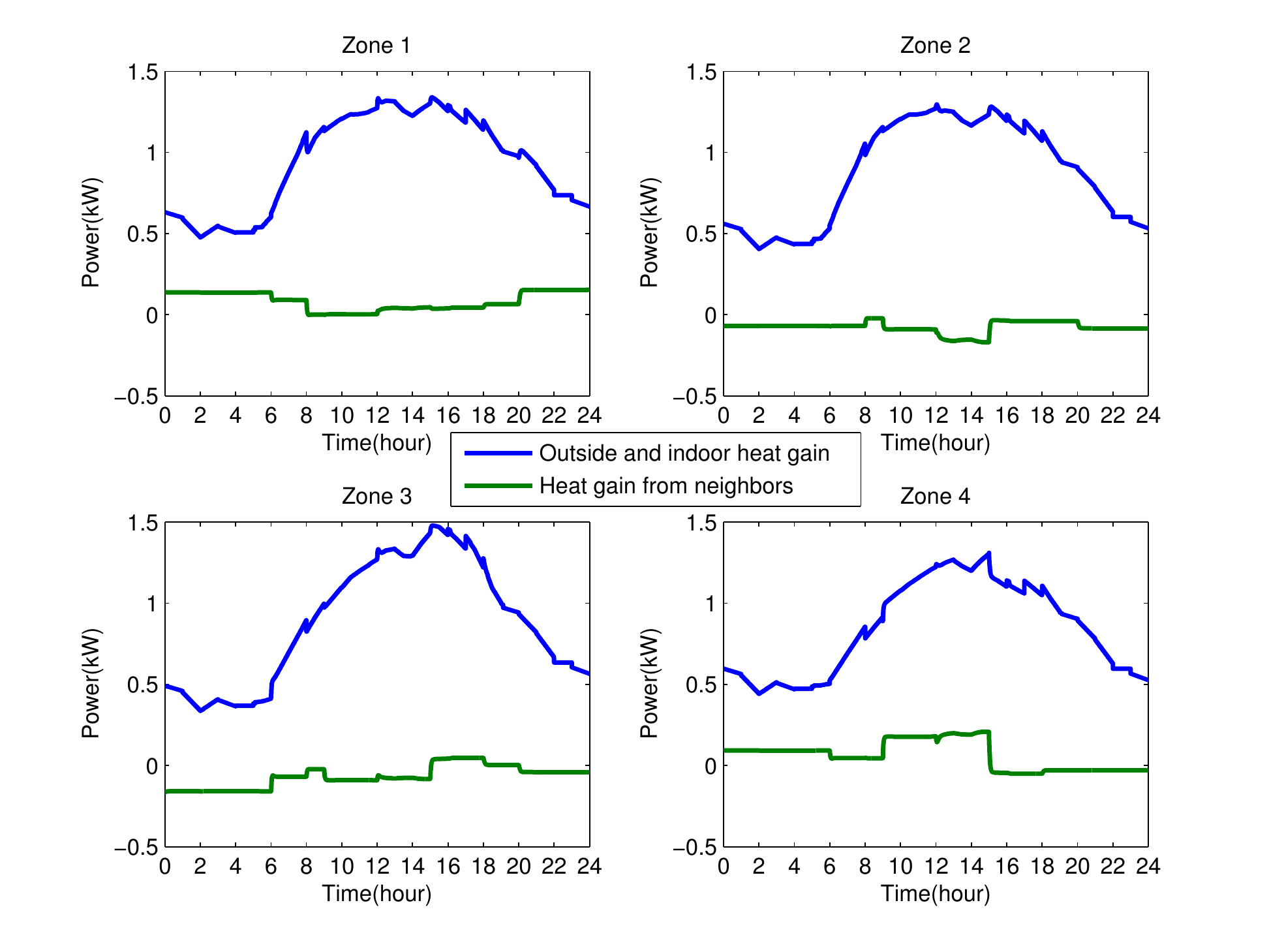}
\caption{Heat gain components in each zone by the accurate model~(\ref{equ:thermalmodel}) under controller~(\ref{equ:controller}b),~(\ref{equ:controller}d)-(\ref{equ:controller}h) and~(\ref{equ:controllerplus}).}
\label{fig:zgainapp}
\end{figure}

\begin{figure}[!t]
\centering
\includegraphics[width=0.5\textwidth]{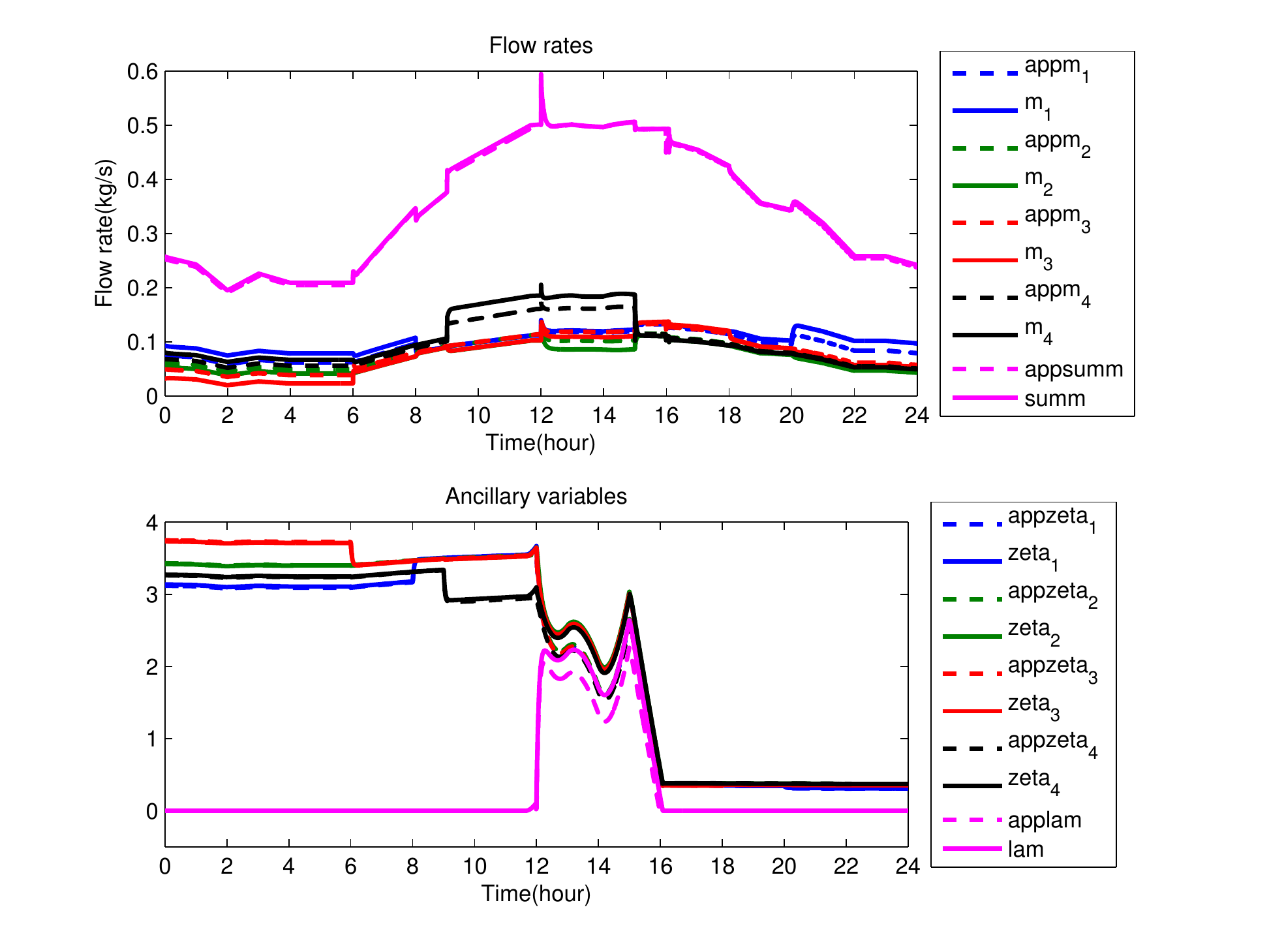}
\caption{Air flow rates in each zone and ancillary variables under controller~(\ref{equ:controller}b),~(\ref{equ:controller}d)-(\ref{equ:controller}h) and~(\ref{equ:controllerplus}): curves labeled with ``app'' indicate the case of using the approximate model~(\ref{equ:thermalmodelapp}).}
\label{fig:ratedualapp}
\end{figure}

The simulation result of the first scenario is illustrated in Figures~\ref{fig:tempapp}-\ref{fig:ratedualapp}, in which we changed $w=1$p.u. to $w=0.1$p.u. at $12$h. The curves labelled with ``app'' indicate the case of using~(\ref{equ:controller}b),~(\ref{equ:controller}d)-(\ref{equ:controller}h) and~(\ref{equ:controllerplus}) for the approximate model~(\ref{equ:thermalmodelapp}) while the curves labelled without ``app'' are for the accurate model~(\ref{equ:thermalmodel}). It can be seen that the difference between these two cases is very small due to that the heat transfer between neighboring zones is negligible compared with the heat gain from the other (outside and indoor) sources, as shown in Figure~\ref{fig:zgainapp} and explained in Section~\ref{se:approximation} as well. The temperature deviations with respect to their set points before decreasing the weight coefficient $w$ are larger than those thereafter since starting from $12$h, user comfort becomes more important while energy consumption saving becomes less. The total flow rate reaches its maximum from $12$h to $16$h, while does not saturate in other periods. Interestingly, there are always deviations from the real temperatures to their set points (unless $w=0$), because there is a tradeoff between user comfort and energy saving when using the proposed controller. All $\zeta_i$ are always positive, indicating that the convex relaxation procedure proposed in Section~\ref{se:approximation} always ensures tightness.

\begin{figure}[!t]
\centering
\includegraphics[width=0.5\textwidth]{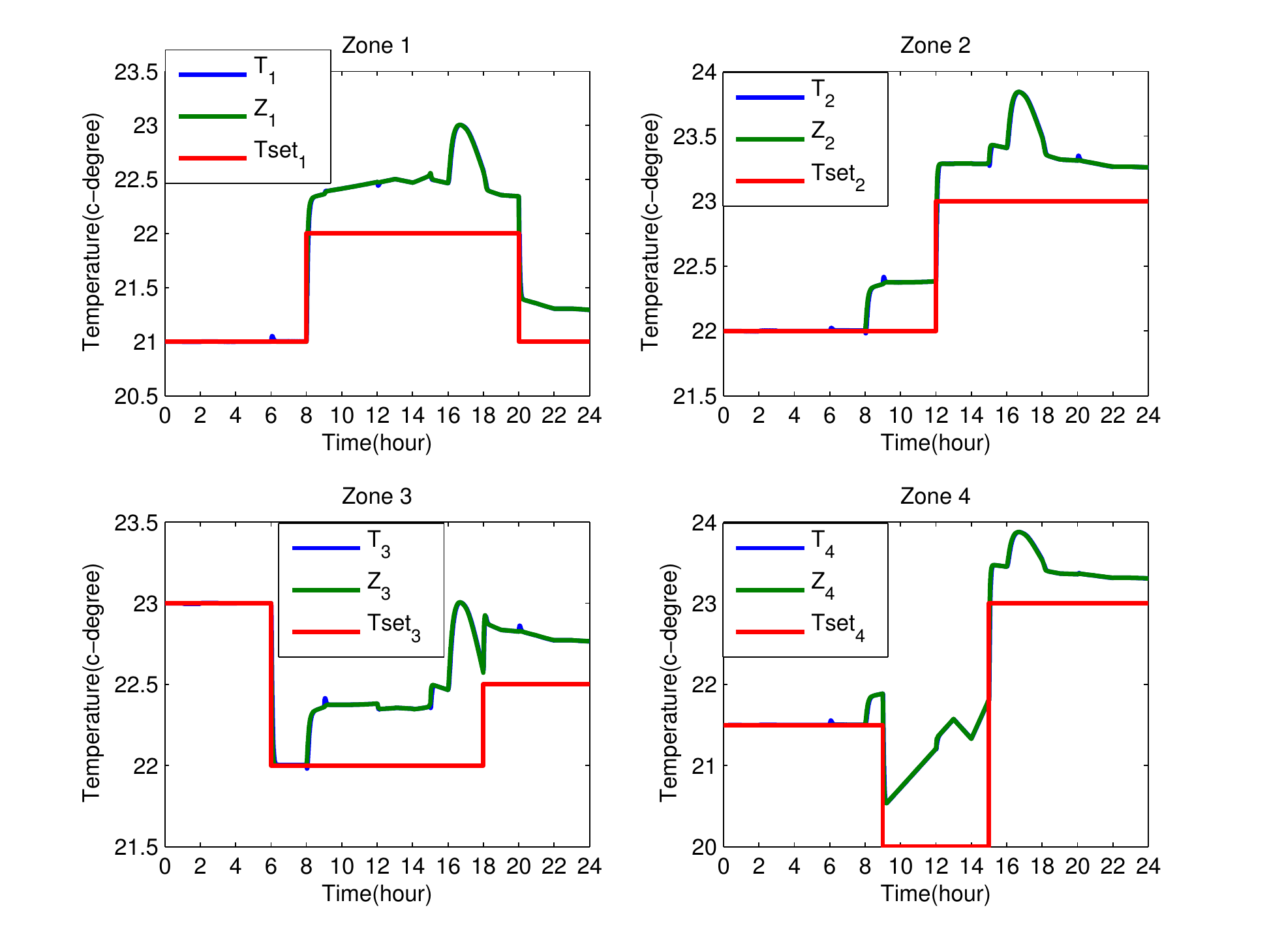}
\caption{Temperatures in each zone under~(\ref{equ:controls=0}b)-(\ref{equ:controls=0}c) and~(\ref{equ:controls=0plus}).}
\label{fig:tempge}
\end{figure}

\begin{figure}[!t]
\centering
\includegraphics[width=0.5\textwidth]{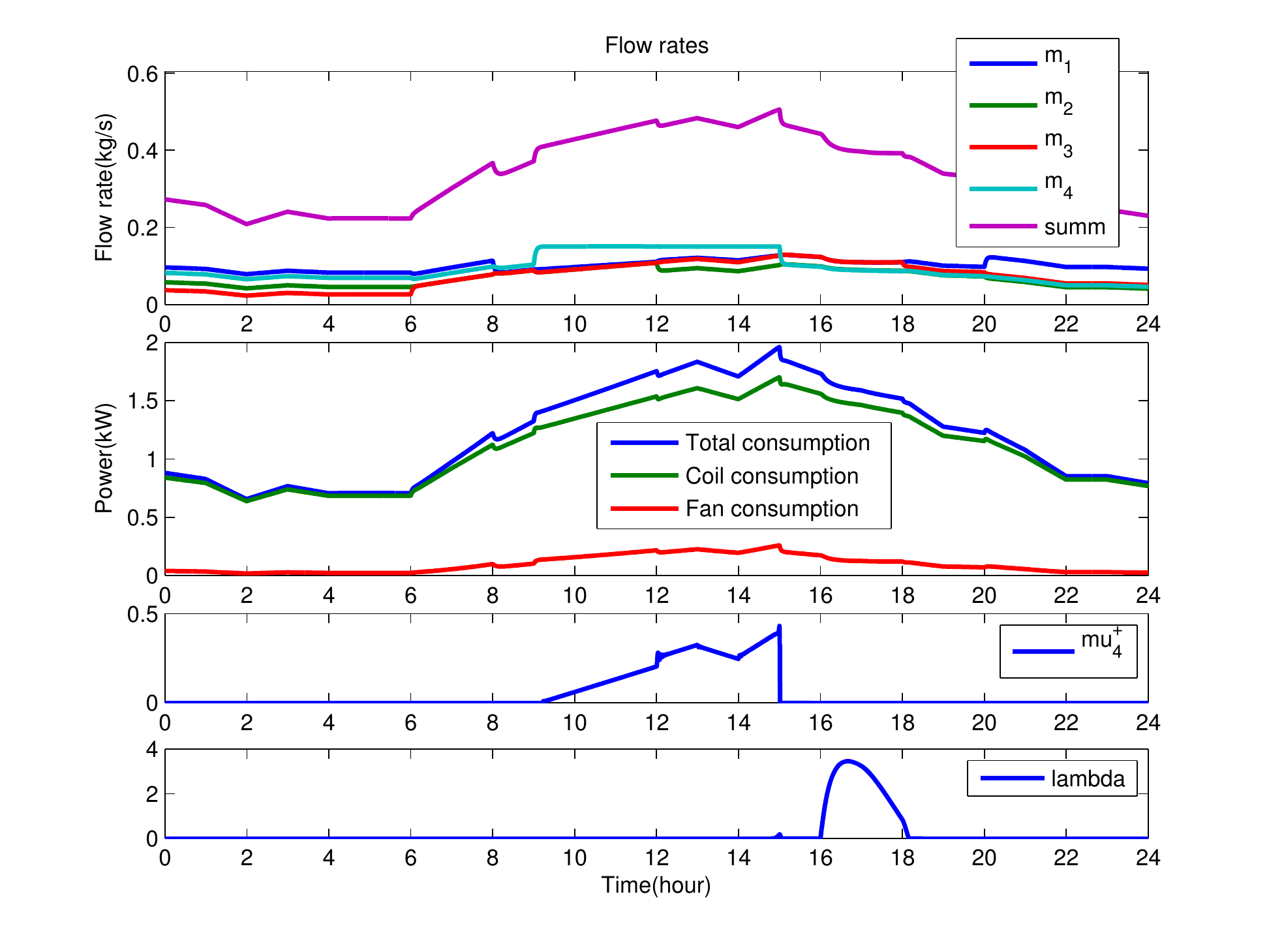}
\caption{Air flow rates, the total consumption and ancillary variables under~(\ref{equ:controls=0}b)-(\ref{equ:controls=0}c) and~(\ref{equ:controls=0plus}).}
\label{fig:ratedualge}
\end{figure}

In the second scenario, we modify the parameters as follows: all comfort ranges are within $\pm 1.8^{\circ}$C of their set points, $[m_4^{min},m_4^{max}]=[0.01,0.15]$kg/s, $\overline{m}=0.5$kg/s before $16$h while $\overline{m}$ decreases to $0.4$kg/s after $16$h, $w=0$ before $8$h and $w=1$p.u. after $8$h (the other parameters remain the same). The simulation result is illustrated in Figures~\ref{fig:tempge}-\ref{fig:ratedualge}. We can see that the curves of the ancillary states $Z_i$ almost coincide with the temperature curves $T_i$ as expected. Before $8$h, the temperatures are exactly the same as their set points since there is no consumption reduction purpose and the total flow rate dose not saturate during this period. After $8$h, deviations appear due to the consideration of energy saving. The total flow rate reaches its maximum around $15$h, and from $16$h to $18$h, thus, the zone temperatures further deviate from their set points (due to the lack of capacity). All these two scenarios inspire us that tuning the weight coefficient $w$ (or equivalently $r_i$, as the optimal solution of~(\ref{equ:opt})/(\ref{equ:optapp}) depends on the ratio $r_i/w, i\in\mathcal{N}$) can balance user comfort and HVAC system energy consumption -- there always exists a tradeoff between user comfort and energy saving.

\section{Conclusion and Future Work}\label{se:conclusion}
This paper presents decentralized/distributed control frameworks on real-time temperature regulation for HVAC systems in energy efficient buildings. The proposed controllers adjust air flow rates in each zone, which balances user comfort and energy saving. Moreover, they can automatically adapt to changes of disturbances such as the outdoor temperature and indoor heat gains, without measuring or predicting those values. Also, the implementation of the controllers are simple.

Future work includes: investigating, e.g., the $H_2$ and $H_{\infty}$ performances of the controlled systems; considering the fact of reheating by each VAV box, and natural ventilation scheduled by users (these will be characterized as extra exogenous inputs to the thermal dynamics); studying the interaction between the controlled building HVAC systems and power grids/microgrids; last but not least, extending the control design frameworks to other heating and cooling systems, for instance, ground source heat pump systems.

\bibliographystyle{IEEEtran}
\bibliography{zxbb}

\begin{thebibliography}{10}
\providecommand{\url}[1]{#1}
\csname url@samestyle\endcsname
\providecommand{\newblock}{\relax}
\providecommand{\bibinfo}[2]{#2}
\providecommand{\BIBentrySTDinterwordspacing}{\spaceskip=0pt\relax}
\providecommand{\BIBentryALTinterwordstretchfactor}{4}
\providecommand{\BIBentryALTinterwordspacing}{\spaceskip=\fontdimen2\font plus
\BIBentryALTinterwordstretchfactor\fontdimen3\font minus
  \fontdimen4\font\relax}
\providecommand{\BIBforeignlanguage}[2]{{%
\expandafter\ifx\csname l@#1\endcsname\relax
\typeout{** WARNING: IEEEtran.bst: No hyphenation pattern has been}%
\typeout{** loaded for the language `#1'. Using the pattern for}%
\typeout{** the default language instead.}%
\else
\language=\csname l@#1\endcsname
\fi
#2}}
\providecommand{\BIBdecl}{\relax}
\BIBdecl

\bibitem{UNEP09}
{UNEP Sustainable Buildings \& Climate Initiative}, ``Buildings and climate
  change: {S}ummary for decision-makers,'' \emph{Paris CEDEX 09, France:
  Sustainable United Nations}, 2009.

\bibitem{USE12}
{U.S. Energy Information Administration Office of Integrated and International
  Energy Analysis, U.S. Department of Energy}, ``Annual energy outlook 2012,''
  \emph{DOE/EIA-0383}, 2012.

\bibitem{NaiR11i}
D.~S. Naidu and C.~G. Rieger, ``Advanced control strategies for heating,
  ventilation, air-conditioning, and refrigeration systems -- an overview:
  {P}art {I}: {H}ard control,'' \emph{HVAC\&R Research}, vol.~17, no.~1, pp.
  2--21, 2011.

\bibitem{NaiR11ii}
------, ``Advanced control strategies for heating, ventilation,
  air-conditioning, and refrigeration systems -- an overview: {P}art {II}:
  {S}oft and fusion control,'' \emph{HVAC\&R Research}, vol.~17, no.~2, pp.
  144--158, 2011.

\bibitem{OldPJ12}
F.~Oldewurtel, A.~Parisio, C.~N. Jones, D.~Gyalistras, M.~Gwerder, V.~Stauch,
  B.~Lehmann, and M.~Morari, ``Use of model predictive control and weather
  forecasts for energy efficient building climate control,'' \emph{Energy and
  Buildings}, vol.~45, pp. 15--27, 2012.

\bibitem{AswMT12}
A.~Aswani, N.~Master, J.~Taneja, D.~Culler, and C.~Tomlin, ``Reducing transient
  and steady state electricity consumption in {HVAC} using learning-based
  model-predictive control,'' \emph{Proc. IEEE}, vol. 100, no.~1, pp. 240--253,
  2012.

\bibitem{MaKDB12}
Y.~Ma, A.~Kelman, A.~Daly, and F.~Borrelli, ``Predictive control for energy
  efficient buildings with thermal storage: modeling, stimulation, and
  experiments,'' \emph{IEEE Control System Magazine}, vol.~32, no.~1, pp.
  44--64, 2012.

\bibitem{Far10}
H.~Farhangi, ``The path of the smart grid,'' \emph{IEEE Power Energy Magazine},
  vol.~8, no.~1, pp. 18--28, 2010.

\bibitem{ZhaP15}
X.~Zhang and A.~Papachristodoulou, ``A real-time control framework for smart
  power networks: {D}esign methodology and stability,'' \emph{Automatica},
  vol.~58, pp. 43--50, 2015.

\bibitem{ZhaKMP15}
X.~Zhang, R.~Kang, M.~McCulloch, and A.~Papachristodoulou, ``Real-time active
  and reactive power regulation in power systems with tap-changing transformers
  and controllable loads,'' \emph{Sustainable Energy, Grids and Networks},
  vol.~5, pp. 27--38, 2016.

\bibitem{ShiLXCG14}
W.~Shi, N.~Li, X.~Xie, C.-C. Chu, and R.~Gadh, ``Optimal residential demand
  response in distribution network,'' \emph{IEEE Journal on Selected Areas in
  Communications}, vol.~32, no.~7, pp. 1441--1450, 2014.

\bibitem{ShiLC17}
W.~Shi, N.~Li, C.~C. Chu, and R.~Gadh, ``Real-time energy management in
  microgrids,'' \emph{IEEE Transactions on Smart Grid}, vol.~8, no.~1, pp.
  228--238, 2017.

\bibitem{JadLM03}
A.~Jadbabaie, J.~Lin, and A.~S. Morse, ``Coordination of groups of mobile
  autonomous agents using nearest neighbor rules,'' \emph{IEEE Transactions on
  Automatic Control}, vol.~48, no.~6, pp. 988--1001, 2003.

\bibitem{Wan10}
F.-Y. Wang, ``Parallel control and management for intelligent transportation
  systems: {C}oncepts, architectures, and applications,'' \emph{IEEE
  Transactions on Intelligent Transportation Systems}, vol.~11, no.~3, pp.
  630--638, 2010.

\bibitem{MorBDB10}
P.~D. Moro\c{s}an, R.~Bourdais, D.~Dumur, and J.~Buisson, ``Building
  temperature regulation using a distributed model predictive control,''
  \emph{Energy and Buildings}, vol.~42, no.~9, pp. 1445--1452, 2010.

\bibitem{MaAB11}
Y.~Ma, G.~Anderson, and F.~Borrelli, ``A distributed predictive control
  approach to building temperature regulation,'' in \emph{Proc. of 2011
  American Control Conference}, 2011, pp. 2089--2094.

\bibitem{HaoLKS15}
H.~Hao, J.~Lian, K.~Kalsi, and J.~Stoustrup, ``Distributed flexibility
  characterization and resource allocation for multi-zone commercial buildings
  in the smart grid,'' in \emph{Proc. of 54th IEEE Conference on Decision and
  Control}, 2015, pp. 3161--3168.

\bibitem{DenBM12}
K.~Deng, P.~Barooah, and P.~G. Mehta, ``Mean-field control for energy efficient
  buildings,'' in \emph{Proc. of 2012 American Control Conference}, 2012, pp.
  3044--3049.

\bibitem{Sug05}
S.~C. Sugarman, \emph{HVAC Fundamentals}.\hskip 1em plus 0.5em minus
  0.4em\relax CRC Press, 2005.

\bibitem{WanWX07}
G.~Wang, Z.~Wang, K.~Xu, and M.~Liu, ``Air handling unit supply air temperature
  optimization during economizer cycles,'' in \emph{Proc. of International
  Conference for Enhanced Building Operations}, 2007.

\bibitem{LinMB12}
Y.~Lin, T.~Middelkoop, and P.~Barooah, ``Issues in identification of
  control-oriented thermal models of zones in multi-zone buildings,'' in
  \emph{Proc. of 51st IEEE Conference on Decision and Control}, 2012, pp.
  6932--6937.

\bibitem{JimMA08}
M.~J. Jim\'{e}nez, H.~Madsen, and K.~K. Andersen, ``Identification of the main
  thermal characteristics of building components using {MATLAB},''
  \emph{Building and Environment}, vol.~43, no.~2, pp. 170--180, 2008.

\bibitem{BacM11}
P.~Bacher and H.~Madsen, ``Identifying suitable models for the heat dynamics of
  buildings,'' \emph{Building and Environment}, vol.~47, no.~7, pp.
  1511--1522, 2011.

\bibitem{HaoLK14}
H.~Hao, Y.~Lin, A.~S. Kowli, P.~Barooah, and S.~Meyn, ``Ancillary service to
  the grid through control of fans in commercial building {HVAC} systems,''
  \emph{IEEE Transactions on Smart Grid}, vol.~5, no.~4, pp. 2066--2074,
  2014.

\bibitem{LinBMM15}
Y.~Lin, P.~Barooah, S.~Meyn, and T.~Middelkoop, ``Experimental evaluation of
  frequency regulation from commercial building {HVAC} system,'' \emph{IEEE
  Transactions on Smart Grid}, vol.~6, no.~2, pp. 776--783, 2015.

\bibitem{BoyV04}
S.~Boyd and L.~Vandenberghe, \emph{Convex Optimization}.\hskip 1em plus 0.5em
  minus 0.4em\relax Cambridge University Press, 2004.

\bibitem{FeiP10}
D.~Feijer and F.~Paganini, ``Stability of primal-dual gradient dynamics and
  applications to network optimization,'' \emph{Automatica}, vol.~46, no.~12,
  pp. 1974--1981, 2010.

\bibitem{CheMC16}
A.~Cherukuri, E.~Mallada, and J.~Cort\'{e}s, ``Asymptotic convergence of
  constrained primal-dual dynamics,'' \emph{System and Control Letters},
  vol.~87, pp. 10--15, 2016.

\bibitem{ZhaLP15cdc}
X.~Zhang, A.~Papachristodoulou, and N.~Li, ``Distributed optimal steady-state
  control using reverse- and forward-engineering,'' in \emph{Proc. of 54th IEEE
  Conference on Decision and Control}, 2015, pp. 5257--5264.

\bibitem{AngCL05}
K.~H. Ang, G.~Chong, and Y.~Li, ``{PID} control system analysis, design, and
  technology,'' \emph{IEEE Transactions on Control Systems Technology},
  vol.~13, no.~4, pp. 559--576, 2005.

\bibitem{DerFS11}
M.~Deru, K.~Field, D.~Studer, K.~Benne, B.~Griffith, P.~Torcellini, B.~Liu,
  M.~Halverson, D.~Winiarski, M.~Rosenberg, M.~Yazdanian, J.~Huang, and
  D.~Crawley, ``Department of energy commercial reference building models of
  the national building stock,'' in \emph{Technical Report NREL/TP-5500-46861},
  2011.

\bibitem{LiWB16}
X.~Li, J.~Wen, and E.~W. Bai, ``Developing a whole building cooling energy
  forecasting model for on-line operation optimization using proactive system
  identification,'' \emph{Applied Energy}, vol. 164, pp. 69--88, 2016.

\end{thebibliography}
\bibliographystyle{choosenstyle}

%\begin{IEEEbiography}
%%[{\includegraphics[width=1in,height=1.25in,clip,keepaspectratio]{xuan}}]
%{Xuan Zhang} received the B.Eng. degree in Control Engineering from Tsinghua University, Beijing, China, in 2011. In 2015, he completed his Ph.D. programme in Department of Engineering Science at the University of Oxford, UK. He is now a Post Doctoral Fellow in both the School of Engineering and Applied Sciences, and Harvard Center for Green Buildings and Cities, at Harvard University, USA. His research interests include the control and optimization for smart power networks, network congestion control, consensus of multi-agent systems, and the stability and robustness analysis of nonlinear dynamical systems.
%\end{IEEEbiography}
%
%
%\begin{IEEEbiography}
%%[{\includegraphics[width=1in,height=1.25in,clip,keepaspectratio]{lina}}]
%{Na Li} received the B.S. degree in mathematics and applied mathematics from Zhejiang University in China and the Ph.D. degree in Control and Dynamical systems from the California Institute of Technology in 2013. She is an Assistant Professor in the School of Engineering and Applied Sciences in Harvard University. She was a postdoctoral associate of the Laboratory for Information and Decision Systems at Massachusetts Institute of Technology. She was a Best Student Paper Award finalist in the 2011 IEEE Conference on Decision and Control. Her research lies in the design, analysis, optimization and control of distributed network systems, with particular applications to power networks and systems biology/physiology.
%\end{IEEEbiography}

\end{spacing}

\end{document}